%% file: neurips_2026.tex
\documentclass{article}

 \usepackage[preprint]{neurips_2026}

\usepackage[utf8]{inputenc} 
\usepackage[T1]{fontenc}    
\usepackage{hyperref}       
\usepackage{url}            
\usepackage{booktabs}       
\usepackage{amsfonts}       
\usepackage{nicefrac}       
\usepackage{microtype}      
\usepackage{xcolor}         
\usepackage{graphicx}

\title{Same Words, Different Judgments:\\How Preferences Vary Across Modalities}

\author{%
  Aaron~Broukhim \\
  Department of Computer Science and Engineering\\
  University of California San Diego\\
  La Jolla, CA 92093 \\
  \texttt{aabroukh@ucsd.edu} \\
  \And
  Nadir~Weibel \\
  Department of Computer Science and Engineering\\
  University of California San Diego\\
  La Jolla, CA 92093 \\
  \texttt{weibel@ucsd.edu} \\
  \And
  Eshin~Jolly \\
  Department of Psychology\\
  University of California San Diego\\
  La Jolla, CA 92093 \\
  \texttt{e3jolly@ucsd.edu} \\
}

\begin{document}

\maketitle

\begin{abstract}
Preference-based reinforcement learning (PbRL) is the dominant framework for aligning AI systems to human preferences. However, evaluation protocols for such data were designed for text and have not been validated for speech. We present the first intraclass correlation coefficient (ICC)-based, controlled cross-modal study of human and synthetic preference annotations, comparing text and audio evaluations of identical semantic content across 100 prompts. We show that achieving \textit{good} (audio) to \textit{near good} (text) agreement within either modality (ICC(2,$k$) $\approx$ .81 audio, .74 text) requires $\sim$9 raters and modalities show marked differences in how people report preferences: audio raters exhibit narrower decision thresholds, reduced length bias, and more user-oriented evaluation criteria, with near-chance cross-modality agreement. We demonstrate that synthetic ratings can be used to effectively predict inter-rater agreement, thus serving as an early signal for stimulus selection and proxy for human annotations. Together, these findings argue that evaluation protocols for audio preference data require modality-specific design rather than direct adaptation from text. \footnote{Code, data, and the audio corpus are available at: \url{https://huggingface.co/datasets/sciminds/same-words-different-judgements}}
\end{abstract}


\section{Introduction}
Speech audio models have seen rapid development and deployment in recent years, yet research on aligning them to human preferences remains sparse~\cite{broukhim2025preference}. Preference-based reinforcement learning (PbRL) offers a path forward, particularly for qualities like naturalness~\cite{huang2025step, zhang2024speechalign}, emotional expressiveness~\cite{gao2025emo}, and conversational appropriateness~\cite{chu2024qwen2}, where vocal delivery may shape perception alongside semantic content.

Current preference alignment research centers on text, following a standard pipeline: collect pairwise preferences, train a reward model, and optimize via RL~\cite{chaudhari2025rlhf}. This approach has revealed biases such as annotators' tendency to prefer longer responses~\cite{shen2023loose}, while largely ignoring additional factors like attention allocation and reading order. Audio by contrast, delivers information at a controlled, recordable pace but requires sequential presentation, making \textit{order} an explicit source of influence on annotations. 

In this work, we empirically investigate: (1)~How reliable are audio preferences? (2)~How does modality affect preference ratings? (3)~Can synthetic ratings substitute or augment human ratings? Our contribution is threefold: (a)~an evaluation protocol for cross-modal preference elicitation (continuous VAS ratings, counterbalanced sequential audio presentation, attention checks adapted for audio); (b)~an empirical characterization of how reliability scales with rater count and how preference judgments shift across modalities with ablations over four TTS voices from three engines and audio-rendered conversation history; and (c)~an annotated corpus of 3{,}113 TTS-converted conversations from PRISM to support future audio preference research. Because most existing PbRL work for speech relies on Text-To-Speech (TTS)-generated audio from existing text datasets, \textit{assumed} to be appropriate for speech use cases~\cite{broukhim2025preference}, we focused our study on TTS-rendered speech of text originally written for reading, and address open questions in Section~\ref{limitations}.

We find audio preference ratings as reliable as text but behaviorally distinct: they introduce recency effects while showing reduced length-bias susceptibility. Using intraclass correlation coefficients (ICC), we present the first reliability analysis of preference annotation in both modalities: a single annotator yields poor reliability, three to four yield moderate reliability, and nine approach good reliability. Cross-modality agreement is near chance, with preference shifts varying by prompt rather than uniformly, suggesting TTS-converted text data may not be appropriate for audio preference pipelines. Synthetic ratings can replace or triage ambiguous response pairs, reducing annotation costs without sacrificing label quality.

\section{Background}
\subsection{Preference-Based Reinforcement Learning}
Preference-Based Reinforcement Learning is a framework in which a reward function is learned from human preferences rather than being explicitly specified. Given pairs (or sets) of trajectory segments or outputs, annotators indicate which they prefer, and these comparisons are used to train a reward model~\cite{christiano2017deep}. This reward model then serves as the optimization objective for a reinforcement learning agent. PbRL has been prominently applied to align Large Language Models with human values through Reinforcement Learning from Human Feedback (RLHF), where preference data is collected by having humans or LLMs compare candidate responses to the same query~\cite{chaudhari2025rlhf}.

Standard PbRL pipelines typically rely on binary pairwise comparisons modeled with Bradley–Terry reward formulations~\cite{christiano2017deep}. While effective, binary labels discard information about preference strength and are susceptible to well-known order and position effects~\cite{basu2019choosing}, which are difficult to diagnose without additional modeling assumptions or counterbalancing. 

Recent work suggests richer scalar feedback yields more informative reward signals than binary comparisons~\cite{wu2023fine}, motivating continuous rating paradigms.

Known biases include length bias~\cite{shen2023loose} (longer responses win regardless of content), position bias~\cite{wang-etal-2024-large-language-models-fair} (first-presented responses are favored), and reward overoptimization~\cite{gao2023scaling}. In audio, these biases are underexplored~\cite{broukhim2025preference}, and many works convert text datasets to audio via TTS~\cite{fang2024llama}, implicitly assuming text-suitable data transfers to speech---an open question.

\subsection{Modality's Effect on Judgments}
Existing work on annotation biases in the speech literature has largely focused on quality assessment, cataloguing biases and proposing mitigations~\cite{zielinski2008some}. Explicit cross-modality comparisons of preference judgments remain rare. Most similar to this work, Cho et al.~\cite{cho2024speechworthy} demonstrated that preferences differ by modality and used this finding to build speech-adapted models, but their analysis did not extend to reliability scaling, order effects, decision thresholds, or synthetic-human alignment---dimensions we address here. 

More broadly, speech processing is strictly linear and imposes higher cognitive load than reading~\cite{cho2024speechworthy}, and users tend to anthropomorphize audio interfaces more readily than text-based ones~\cite{10.1145/3613905.3650818}, suggesting that evaluation criteria may shift across modalities.

\subsection{Reliability and Agreement in Preference Annotations}\label{reliability-review}
Prior work on preference annotation reliability has generally reported only single-rater pairwise agreement---e.g., 73--77\% for text summarization~\cite{stiennon2020learning}, 63\% for helpful and harmless annotations~\cite{bai2022training}, and 60\% for music~\cite{cideron2024musicrl}. While intuitive, these percentages are not directly comparable across studies and, critically, do not address how agreement scales with the number of raters. Intraclass correlation coefficients and Krippendorff's $\alpha$ offer richer characterizations by partitioning variance across raters and stimuli, and by accommodating different measurement scales, respectively~\cite{koo2016guideline}. To our knowledge, no prior work has used these metrics to characterize the number of raters needed to reach specific reliability thresholds in preference annotation for either text or audio.

\subsection{The PRISM Dataset}
PRISM~\cite{kirk2024prism} is a preference dataset of 8,011 live conversations between 1,500 diverse participants and 21 LLMs. Participants rate each response on a 1--100 visual analog scale (VAS) with hidden numeric values to avoid anchoring. Crucially, the person who initiates each conversation is also the one who rates the responses, coupling the evaluative and interactive contexts. We use PRISM's unguided conversations as our source material, adopting its VAS methodology and leveraging its original interactive ratings as our baseline.

\section{Methods}
We converted a subset of PRISM~\cite{kirk2024prism} to spoken audio using Kokoro~\cite{hexgrad_kokoro_82m} (with Chatterbox~\cite{chatterboxtts2025} and Qwen3-TTS~\cite{hu2026qwen3} in a follow-up voice ablation) and collected matched text-to-text and audio-to-audio comparisons. Audio-condition participants also rated audio quality to control for fidelity confounds. This study was approved as minimal-risk research by the University of California San Diego Institutional Review Board (IRB).

\subsection{Converting Text to Audio}
We sub-selected "unguided" conversations to maintain a more objective dataset, since other conversations were intended to be controversial. From the 3,113 unguided conversations, we randomly selected 100 interactions. To ensure variability in response lengths, 25\% (25/100) of the interactions were drawn from the top 10th percentile of absolute character length differences between responses (i.e., a minimum difference of 449 characters). The rest (75/100) were drawn from the lower 90th percentile. This yielded 200 audio clips (2 model responses × 100 interactions). Audio clips were also screened so that only innocuous content was presented to participants, removing clips that could be construed as offensive, discriminatory, or emotionally distressing. Together, this allowed us to experimentally isolate the effect of modality and order on preference ratings.

Each interaction was passed through the current state-of-the-art open-source TTS model, Kokoro~\cite{hexgrad_kokoro_82m}, as benchmarked by TTS-Arena~\cite{tts-arena-v2}, to generate audio samples. We used Kokoro’s default parameters, with the voice set to \texttt{af\_heart} selected for its neutral, clear delivery and kept this consistent across all samples. Follow-up ablation vary voice by using one of Kokoro's additional voices \texttt{am\_michael}, Qwen-3-TTS~\cite{hu2026qwen3} Serena, and Chatterbox's~\cite{chatterboxtts2025} default voice. We note that PRISM responses were originally written by LLMs for text presentation; spoken delivery of text-optimized content may differ from speech written for spoken delivery. Our findings therefore characterize the common pipeline of TTS-converting text-collected data, not the upper bound of audio-native preference annotation (see Section~\ref{limitations}). All audio snippets were manually reviewed for artifacts or mispronunciations that could negatively impact quality to focus on semantic content. Low audio quality samples were replaced, and replacements were re-evaluated following the same selection criteria. We publicly release documentation, the stimuli audio corpus of all 3,113 conversations, analysis code, and website code on \href{https://huggingface.co/datasets/sciminds/same-words-different-judgements}{huggingface}.

The released corpus contains TTS audio for both responses of all 3,113 unguided conversations; the 100-prompt human study is a length-stratified subset, since collecting $\sim$9 ratings per clip per modality across the full corpus was infeasible. The remaining conversations provide stimuli for replication at scale, matched text--audio pairs for training or evaluating audio reward models, and a base for extending our protocol to new rater populations. We additionally release synthetic ratings over the full corpus.

\subsection{Data Collection}
\subsubsection{Demographic Task}
Prior to preference solicitation, participants completed a demographic questionnaire including: age, gender, race, ethnicity, birthplace, education level, employment status, country of residence, primary language, English proficiency (for non-native speakers), other spoken languages, and current timezone. Participants with no English proficiency and those under 18 were excluded. Participant demographics can be found in Appendix Table \ref{tab:demographics}.

\subsubsection{Preference Tasks}
For each interaction, a history of the conversation up to that point in text was shown to the user. The interface can be seen in Appendix Figures \ref{fig:study_interface_audio} and \ref{fig:study_interface_text}. Participants were then asked to rate the responses from 1 (Terrible) to 100 (Perfect) using a slider that does not present the numerical value to avoid anchoring, as done in the original PRISM paper~\cite{kirk2024prism}. Each participant rated 20 interactions presented in a balanced block random order with an attention check at the beginning (21 total). The attention check paired a coherent response with a clearly nonsensical one; participants who preferred the nonsensical response were excluded.

Participants were randomly assigned to either \emph{text} or \emph{audio} response modality to prevent cross-condition contamination. Those in the audio condition also rated audio quality using the same 1-100 scale. Audio quality ratings were collected to monitor for fidelity-related confounds in the response quality judgments. As expected, perceived audio quality was a positive predictor of response ratings within stimuli (see Appendix Section~\ref{audio-quality-robustness}), but controlling for it did not change the order, length, or trial-number effects reported in the main text. We therefore do not include audio-quality ratings as a co-variate in the main models. 

Binary preference data, with the option to tie, were presented to respondents at the end. An optional text box asked raters why they preferred one response over the other for qualitative analysis. We presented both text and audio responses sequentially to explicitly compare order effects across modality.

Audio responses required full playback on the first listen---participants could not skip ahead or scrub through the timeline. After the initial playback, participants could replay the audio as many times as needed, ensuring parity with the text condition where responses could be re-read at will. 

Beyond explicit ratings, we collected additional behavioral indicators (decision time, audio deliberation time, replay counts, intermediate rating changes including reversals after listening to both clips), session-level metadata (duration, completion, dropout), and trial order metadata (which response appeared first, trial position 1--20). We also recorded stimulus properties: history character count and turns, response character length, and absolute character-length differences between paired responses (character length is linearly correlated with audio length, so we treat these as interchangeable for analysis purposes).

\subsubsection{Synthetic Ratings}
We collected synthetic ratings (10 per prompt per modality, 1--100 scale) with default API parameters from three independently developed model families: GPT-4o and GPT-4o-Audio-Preview~\cite{islam2024gpt}, Gemini 3.6 Flash~\cite{google2026gemini36flash}, and Qwen 3.5 Omni Flash~\cite{team2026qwen3}. All audio raters are end-to-end audio models or omni models rather than cascade (STT+LLM+TTS) pipelines, which would only evaluate transcription and synthesis accuracy. We report GPT-4o in the main text and replicate all synthetic-rating analyses with the other two families in Appendix~\ref{assitional-synthetic}. These synthetic scores let us test whether AI signals can augment or replace human annotations for Reinforcement Learning from AI Feedback (RLAIF) applications.

\subsection{Follow-up Ablations}
To test whether our findings depend on a single neutral voice and on text-presented context, we ran two between-subjects ablations on 30-prompt subsets using the same protocol (Appendices~\ref{voice-ablation}--\ref{full-audio-ablation}). \textit{Ablation A} held response content fixed and varied the rendering voice across three additional voices from two additional engines, spanning a $1.0$--$1.9\times$ range of $f_0$ variability (Kokoro am\_michael, Chatterbox, Qwen3-TTS Serena; Table~\ref{tab:r8}). \textit{Ablation B} rendered the conversation history as audio, either in a single voice (B1) or in a distinct voice per speaker (B2); rated responses remained TTS-rendered as in the main study. Each ablation is compared against text and audio reference conditions recomputed on the same prompts. 

\subsection{Participants and Exclusions}
We recruited 106 participants (53 per modality) from Prolific's~\cite{prolific2024} US population at \$12/hour. 7/53 (13.2\%) audio and 5/53 (9.4\%) text participants failed the attention check; of the remainder, 4 text participants did not finish the survey but we retained their completed trials, yielding 94 participants and 8--10 ratings per audio clip. Demographics are reported in Appendix Table~\ref{tab:demographics}. Ablation samples (48 voice-ablation raters; 22 and 23 audio-context raters) are reported in Appendices~\ref{voice-ablation}--\ref{full-audio-ablation}.

\section{Results}
\subsection{Cross-Modal Rating Characteristics}
\begin{figure}[h]
    \centering
    \includegraphics[width=0.6\linewidth]{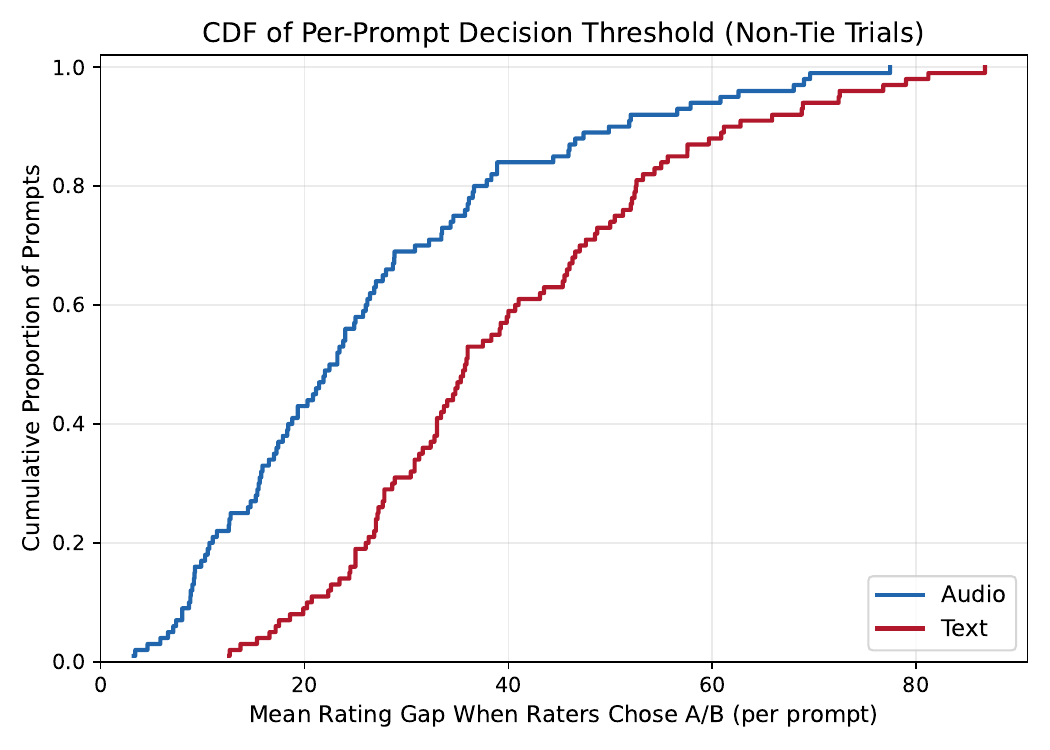}
    \caption{Cumulative distribution of the mean per-prompt rating gap on trials where raters declared a preference (A or B rather than Tie).}
    \label{fig:agreement_cdf}
\end{figure}

\textbf{Decision Threshold.} Raters exhibited larger average rating differences when committing to a winner (not a tie) when presented as text ($M = 41.7$, 95\% CI $[39.5, 43.9]$; $Mdn = 34.0$) compared to audio ($M = 27.9$, 95\% CI $[26.1, 29.8]$; $Mdn = 20.0$; Mann--Whitney $U = 150582$, $p < .001$, $r_{rb} = .303$). Preference ties, while statistically different ($U = 34092$, $p = .022$, $r_{rb} = -.113$), exhibited small magnitude differences (Audio: $M = 3.9$; Text: $M = 3.3$ on a 100-point rating gap). We represent this as a cumulative distribution function in Figure \ref{fig:agreement_cdf}.

\begin{table}[t]
\centering
\caption{Fixed effects from the linear mixed-effects model predicting raw ratings.}
\label{tab:omnibus}
\begin{tabular}{@{}lrcr@{}}
\toprule
Predictor & \multicolumn{1}{c}{$b$} & 95\% CI & \multicolumn{1}{c}{$p$} \\
\midrule
Intercept                          &  74.80  & $[70.72,\ 78.87]$  & $<.001$ \\
Presented first                    & $-2.61$ & $[-4.65,\ -0.58]$  & $.012$  \\
Modality (text)                    & $-6.21$ & $[-11.28,\ -1.14]$ & $.017$  \\
Character length ($z$)             &   3.35  & $[1.10,\ 5.60]$    & $.004$  \\
Trial number ($z$)                 &   0.03  & $[-1.02,\ 1.07]$   & $.960$  \\
Presented first $\times$ Modality  & $-0.51$ & $[-3.42,\ 2.40]$   & $.732$  \\
Modality $\times$ Char.\ length    &   1.53  & $[0.03,\ 3.02]$    & $.045$  \\
Modality $\times$ Trial number     &   0.90  & $[-0.59,\ 2.39]$   & $.236$  \\
\bottomrule
\end{tabular}
\par\smallskip
\footnotesize\textit{Note.} Reference level for modality is audio. CIs are 95\% t-based intervals with Satterthwaite degrees of freedom.
\end{table}

\noindent
\textbf{Stimulus length, order, and fatigue.} Modality moderated the effect of stimulus length ($b = 1.53$, 95\% CI $[0.03, 3.02]$, $p = .045$): longer responses predicted higher ratings in both modalities, but the effect was stronger for text (text: $b = 4.92$, $p < .001$; audio: $b = 3.52$, $p = .004$). A significant recency bias emerged, with second-presented items rated more favorably ($b = -2.61$, $p = .012$), without modality differences ($b = -0.51$, $p = .732$). Ratings remained stable across trials ($b = 0.03$, $p = .960$) with no modality-specific drift ($b = 0.90$, $p = .236$).

\noindent
\textbf{Practice Effects.} A separate model on log-transformed trial duration showed completion times decreased over trials in both modalities (audio: $b = -0.09$, $p < .001$, $\sim$8.6\% reduction per SD of trial number; text: $b = -0.17$, $p < .001$, $\sim$16.0\%), with text participants speeding up more ($b = -0.10$, $p < .001$). Despite these speedups, ratings remained stable, suggesting increased efficiency rather than declining effort.

\subsection{Reliability of Audio Preference Data}

\textbf{Session Characteristics.} Appendix Table \ref{tab:session} summarizes session characteristics. Audio trials took nearly twice as long as text trials ($M = 136$ vs.\ 79 sec; $U = 255.0$, $p < .001$). Attention check failure rates did not differ significantly between conditions (Audio: 13.2\%; Text: 9.4\%; Fisher's exact test, $p = .761$), and participants rated both tasks as similarly difficult ($U = 992.0$, $p = .868$).

\noindent
\textbf{Inter-Rater Reliability (ICC).} We estimated ICC(2,1) and ICC(2,$k$) via linear mixed-effects models with crossed random intercepts for stimuli and participants, with order/trial position as covariates and 95\% CIs from 2,000 bootstrap resamples. Modality differences were tested by permutation (2,000 iterations, pseudo-Audio/Text groups preserving sample sizes). No significant differences were observed between ICC values for audio vs text (ICC(2,1) = .312 vs.\ .241, $p = .094$; ICC(2,$k$) = .806 vs.\ .738, $p = .120$).

\begin{table}[t]
\centering
\caption{Inter-rater reliability metrics by modality.}
\label{tab:reliability}
\begin{tabular}{@{}lccr@{}}
\toprule
Metric & \multicolumn{1}{c}{Audio} & \multicolumn{1}{c}{Text} & \multicolumn{1}{c}{$p$} \\
\midrule
Avg.\ raters per stimulus       & 9.2                          & 8.9                          & ---     \\
ICC(2,1)                        & .312 \,{\scriptsize [.229,\ .378]}    & .241 \,{\scriptsize [.170,\ .299]}    & $.094$  \\
ICC(2,$k$)                      & .806 \,{\scriptsize [.732,\ .848]}    & .738 \,{\scriptsize [.646,\ .792]}    & $.120$  \\
Krippendorff's $\alpha$ (cont.) & .315 \,{\scriptsize [.230,\ .388]}    & .228 \,{\scriptsize [.157,\ .298]}    & $.069$  \\
Krippendorff's $\alpha$ (ord.)  & .157 \,{\scriptsize [.109,\ .206]}    & .217 \,{\scriptsize [.167,\ .268]}    & $.119$  \\
Krippendorff's $\alpha$ (nom.)  & .010 \,{\scriptsize [$-.007$,\ .026]} & .031 \,{\scriptsize [.012,\ .049]}    & $.437$  \\
\bottomrule
\end{tabular}
\par\smallskip
\begin{minipage}{0.7\linewidth}
\footnotesize\raggedright
\footnotesize\textit{Note.} Values are point estimates with 95\% bootstrap CIs from stimulus resampling (2,000 for ICC; 5,000 for $\alpha$). $p$-values from permutation tests of the Audio--Text difference.
\end{minipage}
\end{table}

\noindent\textbf{Inter-Rater Reliability (Krippendorff's $\alpha$).} We computed Krippendorff's $\alpha$ for continuous ratings, ordinal preferences (A/Tie/B relative to presentation order, adjacent disagreements weighted less), and nominal preferences (raw A/Tie/B, all disagreements equal), with 95\% CIs from 5,000 bootstrap resamples and modality tested by permutation (5,000 iterations). Unlike ICC, $\alpha$ was computed without covariate adjustment. No differences were significant (Table~\ref{tab:reliability}). Nominal $\alpha$ was near zero in both conditions, given the counterbalanced presentation order of our experimental design (i.e. agreement would assign opposite A/B labels).

\subsection{Modality Effects on Preferences} \label{modality-effect-pref}
\textbf{Cross-Modality Agreement.} For each prompt, we asked whether audio and text raters agreed on the preferred response. Within each modality, we averaged raw ratings across the $\sim$9 raters per stimulus (justified by the good-to-near-good aggregate reliability, ICC(2,k) = .806 audio, .738 text) and identified the response with the higher mean. We then compared these per-prompt winners across modalities. At a zero-difference threshold, audio and text agreed on the winner for 53\% of prompts (53/100; binomial vs.\ 50\%, $p = .31$). Agreement rose with stricter decisiveness thresholds while the number of decisive pairs dropped (Appendix Figure~\ref{fig:cross-modality-agreement}); including ties-vs.-decisive as disagreements drops agreement until a threshold of $\sim$10.

\noindent
\textbf{Prompt-Specific Modality Effects.} To test whether modality systematically shifted preferences, we computed per-trial preference scores (Clip$_0$ $-$ Clip$_1$ ratings) and fit linear mixed-effects models with modality as a fixed effect and crossed random effects for prompt and participant. The fixed effect of modality was not significant ($\beta = 1.71$, $p = .458$), indicating no overall preference shift between modalities. However, adding random slopes for modality by prompt significantly improved model fit ($\chi^2(2) = 22.1$, $p < .001$; random slope $SD = 12.93$), indicating that modality affected preferences differently across prompts---some prompts elicited stronger preferences in audio, others in text.


\subsection{AI-Human Alignment}\label{human-ai-alignment}
\begin{table}[t]
\centering
\caption{Model estimates for rating comparisons.}
\label{tab:agree}
\begin{tabular}{@{}lrr@{}}
\toprule
\textbf{Comparison} & \multicolumn{1}{c}{\textbf{$\bar{d}$}} & \multicolumn{1}{c}{\textbf{$p$}} \\
\midrule
\multicolumn{3}{@{}l}{\textit{AI-Human Absolute Error}} \\
\quad Audio MAE              &  12.27 & ---     \\
\quad Text MAE               &  14.29 & ---     \\
\quad Text vs.\ Audio (LMM)  &   2.02 & $.035$  \\
\addlinespace
\multicolumn{3}{@{}l}{\textit{PRISM vs.\ External Annotators}} \\
\quad vs.\ AI-Text           & $-10.29$ & $<.001$    \\
\quad vs.\ AI-Audio          & $-17.62$ & $<.001$   \\
\quad vs.\ Human-Text        &   $-0.41$  & $.824$   \\
\quad vs.\ Human-Audio       &   $-7.29$  & $<.001$    \\
\bottomrule
\end{tabular}
\par\smallskip
\begin{minipage}{0.7\linewidth}
\footnotesize\raggedright
\textit{Note.} For AI-Human error, $\bar{d}$ is the mean absolute error (MAE) between AI and human clip-level ratings on a 1--100 scale; the LMM row tests the modality difference. For PRISM comparisons, $\bar{d}$ is the mean signed difference (PRISM $-$ comparison group).
\end{minipage}
\end{table}

\noindent
\textbf{Average AI-Human Rating Error.} For each clip we computed mean AI and human ratings within modality and the absolute error $|$AI$-$Human$|$, then fit a linear mixed-effect model (LMM) with modality as a fixed effect and clip as a random intercept.

\noindent
\textbf{PRISM vs.\ External Annotators.} To test whether the original PRISM ratings, where the person interacting with the model did the rating, differed from external annotators, we conducted paired $t$-tests comparing PRISM ratings to each external group (AI-Text, AI-Audio, Human-Text, Human-Audio) at the clip level (raw score mean). We report our results in Table~\ref{tab:agree}. Human-text ratings aligned most closely with PRISM ratings ($\bar{d} = -0.41$, $p = .824$), while human-audio raters scored responses higher than the original PRISM respondents ($\bar{d} = -7.29$, $p < .001$). AI annotators also rated above PRISM, more so in audio ($\bar{d} = -17.62$) than in text ($\bar{d} = -10.29$).

\textbf{AI Difference Predicts Human Agreement.} For each prompt we computed human ICC(2,1) and the absolute mean difference between AI ratings for the two responses, then regressed ICC $\sim$ AI\_diff $\times$ Modality with Spearman correlations as non-parametric checks (Table~\ref{tab:AI_Pred}). When AI ratings sharply distinguished between responses, humans agreed more with each other in both modalities. Results were robust to removal of influential observations (Cook's D). Both findings replicate with Gemini 3.6 Flash and Qwen 3.5 Omni Flash: AI--human error was lower in audio for all three raters ($\bar{d} = 2.02$--$6.46$; Table~\ref{tab:r1}), and AI differentiation predicted human ICC in both modalities, with no modality interaction, for all three (Table~\ref{tab:r2}).

\begin{table}[h]
\centering
\caption{Regression: AI Differentiation and Human Agreement}
\label{tab:AI_Pred}
\begin{tabular}{lcc}
\toprule
Predictor & $B$ & $p$ \\
\midrule
AI Differentiation & 0.014 & $<$.001 \\
Modality (Text) & $-$0.010 & .76 \\
AI Diff $\times$ Modality & $-$0.003 & .15 \\
\midrule
\multicolumn{3}{l}{\textit{Spearman correlations}} \\
Audio & \multicolumn{2}{c}{$r = .23$, $p = .02$} \\
Text & \multicolumn{2}{c}{$r = .38$, $p < .001$} \\
\bottomrule
\end{tabular}
\par\smallskip\noindent\scriptsize 
DV = per-prompt and modality ICC; $N = 100$\\
\end{table}

\subsection{Robustness to Voice and Audio Context}\label{robustness}
\textbf{The threshold gap persists under acoustically uniform rendering.} Relative to text, decision thresholds remained narrower for both new uniform voices (am\_michael: $+12.6$ raw points, Tukey $p = .036$; Serena: $+12.7$, $p = .028$) and when history was heard in a single voice (B1: $+17.5$, $p < .001$; Table~\ref{tab:r15}). B1 was statistically equivalent to the audio baseline on both threshold metrics by two one-sided tests (TOST; Holm $p = .014$ raw, $p = .024$ scaled; Table~\ref{tab:r10b}). B1 showed anomalously low reliability (ICC(2,1) $= .032$), traceable to a subgroup of non-differentiating raters; excluding the faster half of raters yields a threshold (19.6) matching the audio baseline (19.0; Appendix~\ref{full-audio-ablation}).

\noindent
\textbf{Acoustic variation attenuates the gap.} The most expressive voice (Chatterbox, $1.90\times$ baseline $f_0$ variability) widened the scaled threshold relative to the audio baseline ($+0.225$, unadjusted $p = .011$, Tukey $p = .079$), closing $\sim$61\% of the audio--text gap. Rendering history with a distinct voice per speaker widened thresholds relative to single-voice history (B2 $-$ B1: $+12.9$ raw, Tukey $p = .015$), closing $\sim$68\%. Recency effects did not differ across voices (Arm $\times$ position, $p = .62$), whereas length sensitivity did (Arm $\times$ length, $p = .003$) and was near zero when history was heard (Table~\ref{tab:r11}). Synthetic--human alignment was descriptively strongest for the baseline voice (Table~\ref{tab:r9}).

\section{Discussion}
Using a controlled cross-modal experiment, we provide critical analyses examining the reliability and consistency of human preference ratings across text and audio. While modalities are comparably reliable in aggregate, they produce meaningfully different evaluation dynamics: audio raters operated with narrower decision thresholds, showed reduced sensitivity to response length, and described their preferences in user-oriented rather than content-depth terms. 

\subsection{Audio Preference Reliability}\label{audio-pref-rel}
\begin{figure}[h]
  \centering
  \includegraphics[width=0.6\columnwidth]{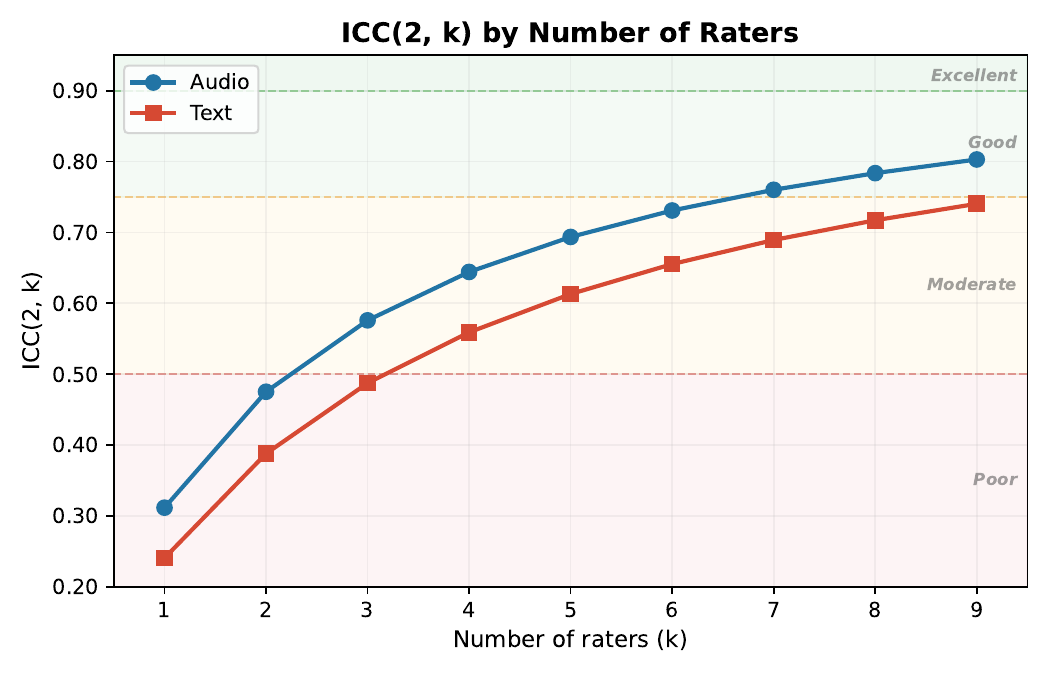}
  \caption{ICC(2,\,\textit{k}) as a function of the number of raters, derived from variance components of a crossed random-effects model. Shaded bands indicate conservative interpretation thresholds per~\cite{koo2016guideline}. The observed average number of raters per stimulus was $k \approx 9.2$ (audio) and $k \approx 8.9$ (text).}
  \label{fig:icc-by-k}
\end{figure}

\textbf{Moderate reliability requires 3-4 raters per stimulus.} To our knowledge, our work is the first to systematically characterize the number of raters needed to reach specific reliability thresholds in the preference annotation literature. As reviewed in Section~\ref{tab:reliability}, prior work has reported only single-rater pairwise agreement and not characterized how reliability scales with rater count. In line with Koo et al.'s~\cite{koo2016guideline} conservative ICC thresholds (Figure \ref{fig:icc-by-k}), agreement across single raters ($k=1$) is \textit{poor} ICC(2,1) = .312 audio, .241 text. At least 3 to 4 raters are required to achieve \textit{moderate} consistency, while \textit{good} agreement is observed with ${\sim}9$ audio raters ICC(2,\textit{k}) = .806 and near good ICC(2,\textit{k}) = .738 for text, with diminishing returns beyond this. Practically, \textit{moderate} reliability preference data demands a \textit{3$\times$ increase} in annotation cost, while reaching \textit{good} agreement demands a \textit{9$\times$ increase} -- which may or may not be justified depending on the requirements of the downstream task.

\noindent 
\textbf{Assuming fixed (text) decision thresholds will misclassify audio preferences.} The compressed audio margin (Figure~\ref{fig:agreement_cdf}) matters for pipelines that convert continuous ratings into binary labels: a fixed binarization threshold calibrated on text may misclassify audio pairs as ties or reverse preferences, inflating label noise for reward model training. We recommend calibrating thresholds per modality or, where possible, retaining the continuous signal to preserve preference strength.

\noindent 
\textbf{Recency Bias Must Be Handled Carefully.} Text preferences are often presented side-by-side to mitigate order effects~\cite{basu2019choosing}, but this is impossible for long audio. Our results confirm a strong recency bias that did not differ by modality, indicating it is a general problem of sequential presentation rather than audio-specific. Counterbalanced presentation order can mitigate this at the aggregate level but inflates individual-level disagreement, likely explaining our near-zero nominal Krippendorff's $\alpha$. Re-listening to the first clip after hearing both has also been proposed~\cite{zielinski2008some}; in our study, 214/920 audio trials (23.3\%) involved voluntary re-listens to the first clip after the second. Binary ratings are especially susceptible to recency since order cannot be adjusted post-hoc, so we recommend continuous ratings for audio preference collection.

\noindent 
\textbf{Longer Rating Sessions Are More Efficient Without Loss of Data Quality.} Despite audio trials taking nearly twice as long, participants rated both conditions as equally difficult, suggesting the additional time reflects audio's pacing rather than higher cognitive demand. Ratings remained stable across trials even as completion times decreased, indicating practice-driven efficiency rather than declining effort. While time-on-task effects are well documented in crowdsourcing~\cite{see1995meta, mason2012conducting}, no prior work has examined them for preference annotation specifically. Our results suggest sessions of up to one hour are viable for both modalities, though future work should identify the degradation boundary.

\subsection{Cross-Modality Considerations}
\textbf{Text and Audio Converge Only When Preferences are Strong.} Cross-modality agreement was only 53\% at a zero threshold, rising with stricter decisiveness (Figure~\ref{fig:cross-modality-agreement}). The divergence is prompt-specific rather than a uniform directional shift: the fixed effect of modality was non-significant, but random slopes for modality-by-prompt substantially improved model fit, implying \textit{modality can swing preferences by nearly half a standard deviation in either direction} depending on the prompt. 

Our TF-IDF analysis of written justifications offers one interpretation: audio raters more frequently referenced ``user'' and ``help'', while text raters emphasized ``detail'' and ``response'' (Appendix~\ref{qualitative}). This is consistent with audio evaluation promoting a holistic, user-oriented frame---\textit{does this response serve the person?}---while text evaluation encourages analytic attention to content depth, aligning with evidence that users anthropomorphize audio interfaces more readily than text-based ones~\cite{10.1145/3613905.3650818}. Another interpretation is the relevance of context, as prompts that are interpretable only given the preceding turn (40/100) flipped winners across modalities far more often than self-contained prompts (68\% vs.\ 35\%), indicating that modality may moderate the importance of a between turn coherency.

\noindent
\textbf{Audio Preferences are Less Susceptible to Length Bias.} Length bias is a well-documented confound in text preference annotation~\cite{singhal2023long}, and we replicate it: longer text responses predicted a $\sim$40\% larger rating increase than longer audio responses (interaction $p = .045$). Audio's fixed pacing likely constrains length as a heuristic since listeners cannot skim or visually gauge response size. This extends Cho et al.~\cite{cho2024speechworthy}, who showed verbosity is penalized more heavily in audio than text---though their best model still generated longer-than-baseline responses while being preferred, indicating audio length bias is moderated rather than reversed. One speculative use case: audio-derived labels could serve as a complementary signal to debias text-trained reward models by upweighting audio scores, at the cost of $\sim$2$\times$ longer annotation per trial---whether this trade-off pays off remains open.

\noindent
\textbf{Acoustic Variation Moderates, but Does Not Explain, Narrower Audio Thresholds.} One account of narrower audio thresholds is that a flat, neutral voice compresses perceived differences between responses. Our ablations partly support it: the most expressive voice and speaker-differentiated context both widened thresholds toward text. Because Chatterbox differs from the baseline in engine, speaker, and speaking rate as well as pitch variability, we treat expressivity as a leading candidate moderator rather than an isolated cause. Yet narrowing persisted in every acoustically uniform condition, including a different engine at near-baseline expressivity and history that was heard rather than read. The gap therefore appears to have two components, one tied to acoustically uniform rendering and one to modality itself. Acoustic properties such as expressivity, speaker differentiation, and pitch register should accordingly be measured and reported alongside content as evaluation factors, rather than held fixed and ignored.

\subsection{Synthetic Rating Applications}
\textbf{Synthetic ratings are an early identifier of high/low consistency prompts.} When AI ratings distinguished between two responses, human raters also agreed more with each other, in both modalities. This finding has immediate practical value: \textit{synthetic ratings can serve as a pre-screening signal to triage prompts before human annotation}. Prompts where AI models strongly differentiate likely need fewer human raters, while ambiguous AI ratings may warrant larger annotator pools to resolve genuine disagreement. Such adaptive sampling could substantially reduce annotation costs without sacrificing label quality. To our knowledge, this relationship between AI discriminability and human inter-rater reliability has not been previously reported, in part because ICC is rarely computed in the preference annotation literature. Future work may use our ICC data to validate other LLMs.

\noindent 
\textbf{\textit{Modality}-dependent AI Ratings Are Viable Human Proxies.} AI--human absolute error was significantly higher for text than audio. While modest on a 100-point scale, this gap is notable given audio's smaller decision thresholds---a 2.0-6.5 point discrepancy represents a proportionally larger share of the signal that distinguishes preferred from non-preferred audio responses. Prior work reports $\sim$80\% GPT-4o-human agreement for text~\cite{zheng2023judging}, and RLAIF can substitute AI for human labels in text with minimal performance loss~\cite{10.5555/3692070.3693141}. Our results extend this to audio, suggesting synthetic ratings are viable proxies in both modalities, though tighter AI-human alignment in audio warrants further investigation into whether end-to-end audio models capture signals unavailable to text pipelines. 

\subsection{Recommendations for Audio Preference Annotation}
Our results suggest four practical guidelines. (1) Calibrate binarization thresholds per modality and rendering condition. (2) Collect continuous rather than binary ratings for audio, since sequential presentation induces recency effects that binary labels cannot absorb post hoc. (3) Counterbalance presentation order to counteract order effects. (4) Triage annotation budgets with synthetic ratings: in a resampling analysis (Table~\ref{tab:r4}), a single human rating matched the full-panel winner on 76--78\% of prompts in the widest AI-gap band versus 49--52\% in the narrowest, and wide-gap prompts stabilized with $\sim$2 raters rather than $\sim$3. A simple policy allocates 2 raters to prompts whose AI rating gap exceeds 9 points (audio) or 23 points (text) and 3 otherwise, saving $\sim$33\% of annotation cost on triaged prompts. 


\section{Limitations}\label{limitations}
Converting text-based interactions to audio via TTS is currently the standard approach for generating speech preference data. As such our findings speak to this common use-case, and allowed us to deliberately control for confounds by: removing potentially harmful or offensive content, excluding ``safety'' and ``refusal'' scenarios, and a US-based, English-speaking sample of participants. However, TTS-rendered speech lacks paralinguistic nuances of natural speech---emotional prosody, hesitation, background noise---that may influence real-world preference judgments. Our claims are therefore scoped to TTS-rendered speech under the acoustic conditions tested, with expressivity identified as a candidate moderator rather than an isolated cause. Speaker characteristics beyond those varied here (accent, culture, naturalistic recordings) and non-innocuous or toxic semantic content are also likely to influence real-world preferences. Preference differences between non-Western populations and non-native English speakers are largely unexplored. 

\section{Conclusion}
Despite the rapid development and deployment of speech audio models in recent years, differences between audio and text modalities has been understudied. We find that modality meaningfully shapes evaluation behavior: preferences vary by prompt rather than shifting uniformly, audio judgments show narrower decision thresholds and reduced length sensitivity, and synthetic ratings align with human judgments while predicting inter-rater reliability. Our work suggests that Text-derived TTS preferences cannot be \textit{assumed} to generalize to speech, and modality should be treated as a first-class consideration in preference-based RL pipelines. At the same time, future work should systematically include more naturalistic audio factors and diverse speech conditions to further understand the differences between modalities.

\begin{ack}
Funding for this project was supported by the NIH National Library of Medicine's T15 Biomedical Informatics and Data Science Research Training Program (Grant T15LM011271) to A.B. who is affiliated with the Department of Biomedical Informatics at UC San Diego.

We'd like to thank Zachary Novack and Saketh Kasibatla for their feedback and continued support throughout this project. 
\end{ack}

\medskip

{
\small
\bibliographystyle{plain}  
\bibliography{references} 


}
\appendix
\section{Broader Impacts}\label{impacts}
Our methodological findings and released audio corpus aim to support audio-aligned speech systems, an area where alignment research remains underdeveloped relative to text. Three considerations warrant care. First, better audio preference pipelines could in principle be misused to optimize persuasive or manipulative speech systems, though our contribution is methodological rather than a deployable model, making this path indirect. Second, our US-based, predominantly White, predominantly native-English-speaking sample (Appendix~\ref{app:demographics}) means reward signals derived from data collected via our protocol could entrench preferences that systematically devalue non-Western speech patterns, accents, or dialects; we recommend demographically broader rater pools for any downstream deployment. Third, we excluded potentially harmful or offensive content from our stimuli, so our reliability and modality-effect results do not characterize safety or refusal scenarios and should not be generalized to safety-critical alignment contexts. We flag the latter two caveats in Limitations (Section~\ref{limitations}).

\section{Study Interfaces}
\begin{figure}[h]
  \centering
  \includegraphics[width=\linewidth]{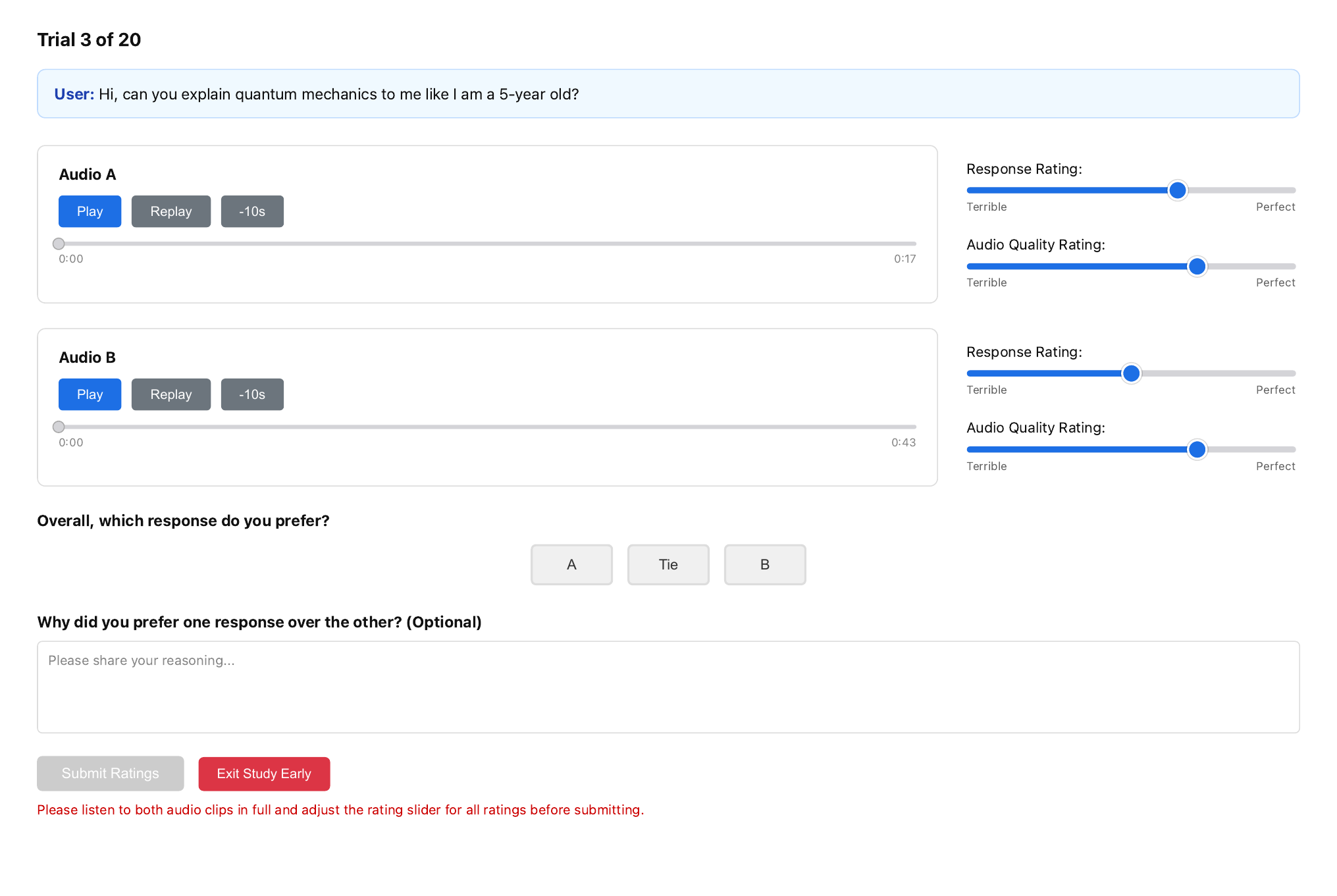}
  \caption{Interface for the Audio Task of preference selection. It includes conversation history, most recent query and two responses. Raters must rate each response from 1-100 on response quality (the content of the answer) and audio quality (the clarity and naturalness of speech). We also collect discrete preferences with the option to tie and optional free-response reasoning justification.}
  \label{fig:study_interface_audio}
\end{figure}

\begin{figure}[h]
  \centering
  \includegraphics[width=\linewidth]{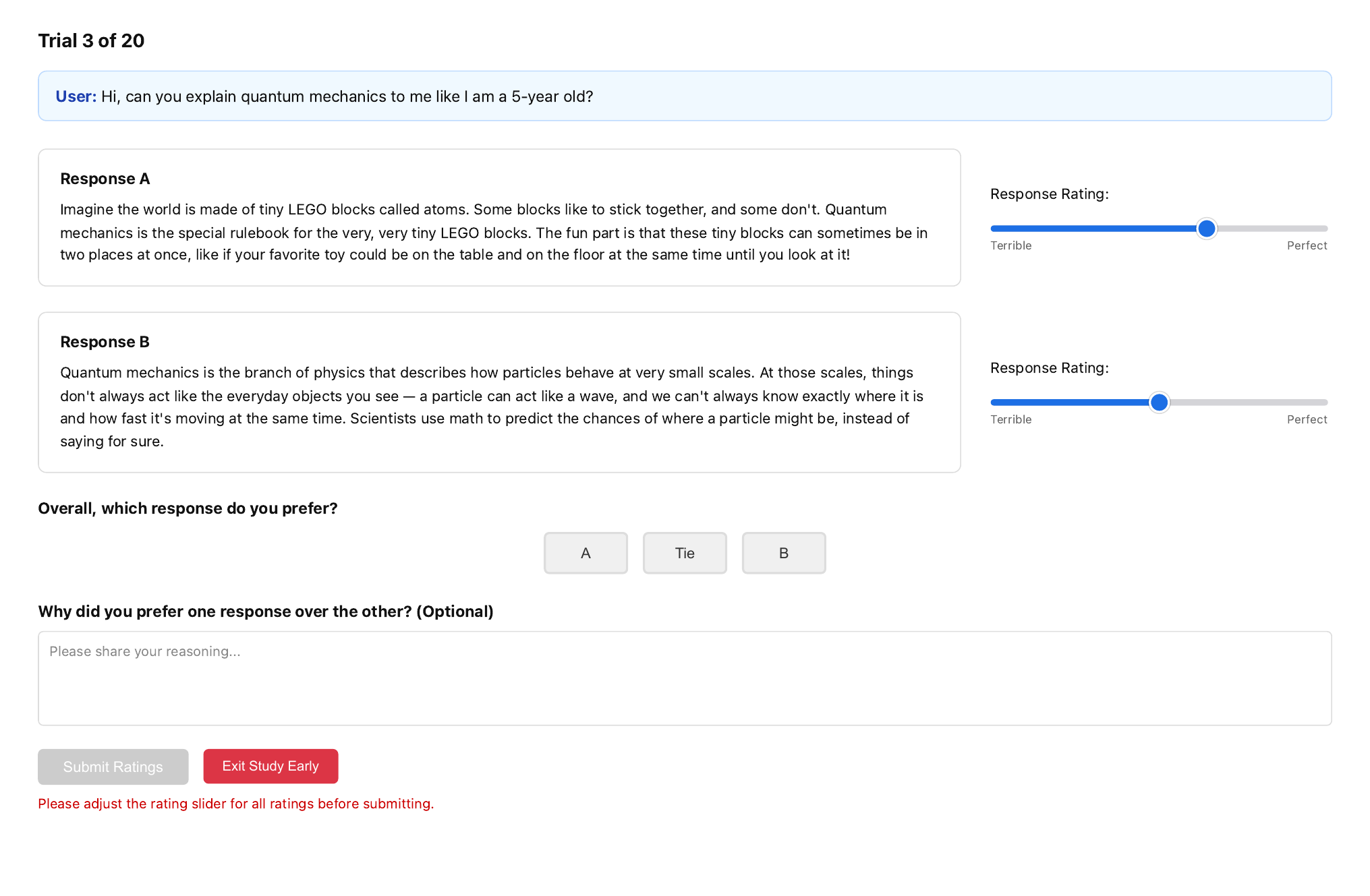}
  \caption{Interface for the Text Task of preference selection. It includes conversation history, most recent query and two responses. Raters must rate each response from 1-100 on response quality (the content of the answer). We also collect discrete preferences with the option to tie and optional free-response reasoning justification.}
  \label{fig:study_interface_text}
\end{figure}

Figures~\ref{fig:study_interface_audio} and~\ref{fig:study_interface_text}
show the rater-facing interfaces for the audio and text conditions,
respectively. The two interfaces were matched on layout, conversation
history presentation, and rating instruments; they differ only in (i)
whether responses were delivered as audio clips with playback controls
or as rendered text, and (ii) the presence of an additional audio-quality
slider in the audio condition. Both conditions used continuous 1--100
visual analog scales with hidden numeric values, followed by a discrete
preference judgment (A / Tie / B) and an optional free-text justification.

\section{Session and Trial Characteristics}
Table~\ref{tab:session} summarizes session- and trial-level descriptive
statistics by modality. Audio sessions took substantially longer than
text sessions both per trial (136 vs.\ 79 seconds) and overall (47 vs.\
27 minutes), reflecting audio's fixed temporal pacing rather than
greater task difficulty: self-reported difficulty ratings were
indistinguishable across conditions ($M = 2.50$ vs.\ 2.45 on a 5-point
scale, $p = .868$). Completion and attention-check failure rates were
also comparable between conditions, suggesting the longer audio sessions
did not disproportionately tax participants or induce dropout. Together,
these characteristics support the interpretation in Section \ref{audio-pref-rel} that audio
annotation imposes a time cost without a corresponding effort cost.


\begin{table}[h]
\centering
\caption{Session and trial characteristics by condition.}
\label{tab:session}
\begin{tabular}{@{}lccr@{}}
\toprule
& \textbf{Text} & \textbf{Audio} & \multicolumn{1}{c}{\textbf{$p$}} \\
\midrule
Completed (\%)         & 92.5         & 96.2          & $.678$\textsuperscript{a}  \\
Session duration, min  & 27.0 (13.3)  & 47.3 (12.4)   & $<.001$\textsuperscript{b} \\
Trial duration, sec    & 78.8 (41.6)  & 136.4 (35.9)  & $<.001$\textsuperscript{b} \\
Difficulty rating      & 2.45 (0.90)  & 2.50 (0.96)   & $.868$\textsuperscript{b}  \\
Failed attention (\%)  & 9.4          & 13.2          & $.761$\textsuperscript{a}  \\
\bottomrule
\end{tabular}
\par\smallskip
\footnotesize\textit{Note.} Values are $M$ ($SD$) unless noted. \textsuperscript{a}Fisher's exact test. \textsuperscript{b}Mann--Whitney $U$.
\end{table}

\section{Cross-Modality Agreement by Decisiveness Threshold}
\begin{figure}[h]
    \includegraphics[width=\textwidth]{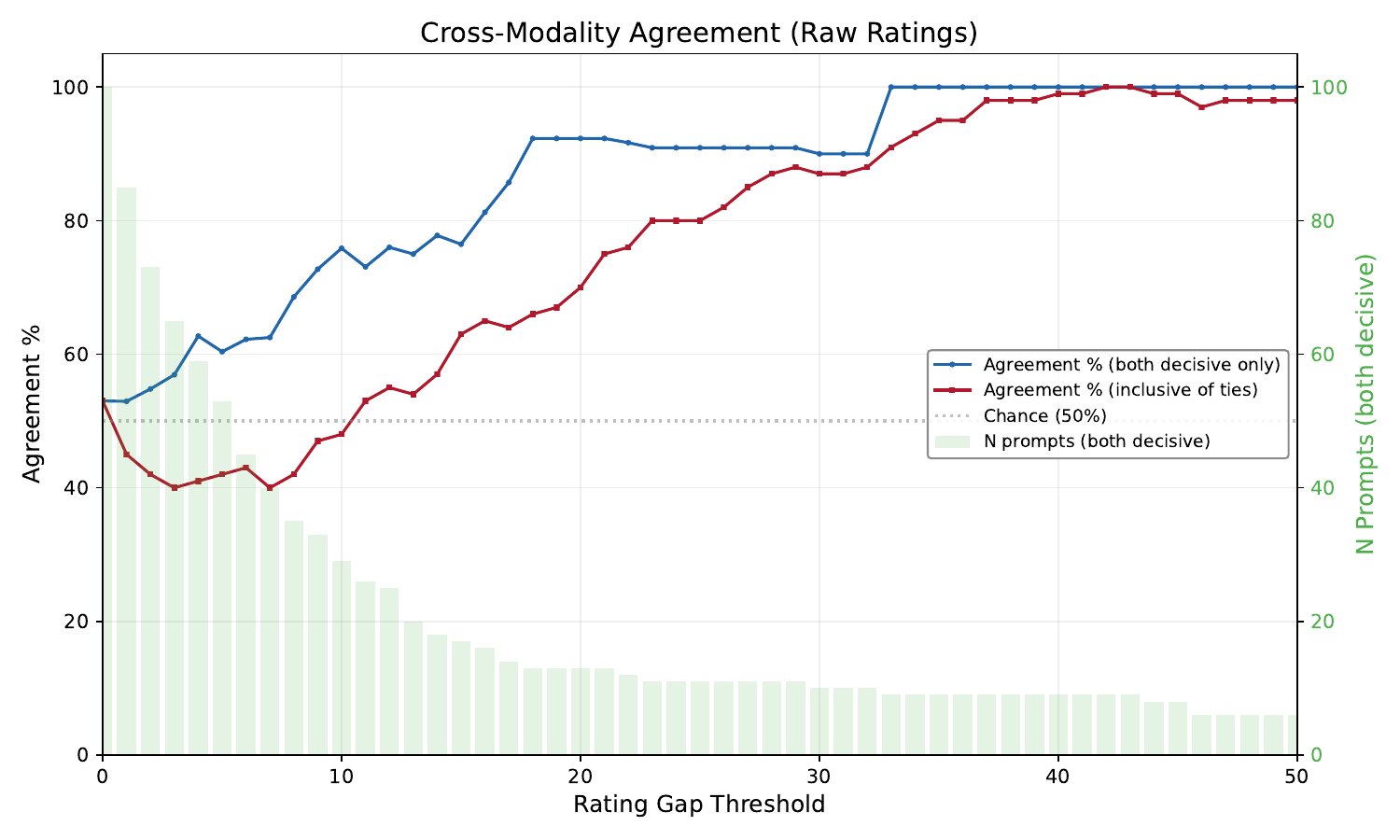}
    \caption{Cross-modality agreement between audio and text evaluations as a function of raw rating gap threshold. The blue line shows agreement percentage when both modalities produce a decisive winner (excluding ties), while the red line shows agreement when tie-versus-decisive outcomes are counted as disagreements. Green bars indicate the number of prompts where both modalities had a decisive winner at each threshold.}
    \label{fig:cross-modality-agreement}
\end{figure}

Figure~\ref{fig:cross-modality-agreement} extends the Section~\ref{modality-effect-pref} analysis by
plotting cross-modality winner agreement as a function of the minimum
rating gap required to count a prompt as ``decisive.'' The blue curve
restricts the comparison to prompts where both modalities produced a
decisive winner at the given threshold; agreement rises monotonically
from 53\% at a zero-difference threshold to substantially higher rates
once only strongly preferred pairs are retained, but the number of
qualifying prompts (green bars) drops accordingly. The red curve treats
tie-versus-decisive mismatches as disagreements, and consequently
declines until the threshold reaches roughly 10 points---the point at
which most prompts in both modalities have settled into clear
preferences. The two curves together indicate that text and audio raters
converge primarily on the prompts where preferences are unambiguous, and
diverge on the ambiguous middle, consistent with the prompt-specific
modality effects reported in Section~\ref{modality-effect-pref}.

\section{Robustness to Perceived Audio Quality}\label{audio-quality-robustness} Because the audio condition introduced a fidelity dimension absent in text, we re-fit the audio-only simple-effects model with $z$-scored audio-quality ratings added as a covariate ($n = 1{,}840$ observations, 46 participants, 200 stimuli). Perceived audio quality was, unsurprisingly, a strong positive predictor of response ratings ($b = 7.77$, 95\% CI $[6.51, 9.03]$, $p < 10^{-31}$): a 1-SD increase in audio-quality rating corresponded to a $\sim$7.8-point increase in response rating. However, the substantive effects from the main text were preserved. The recency bias remained significant ($b = -2.28$, 95\% CI $[-4.11, -0.44]$, $p = .015$; original $b = -2.63$, $p = .006$, $\sim$14\% attenuation). The length effect was essentially unchanged ($b = 3.27$, 95\% CI $[1.05, 5.50]$, $p = .004$; original $b = 3.53$, $p = .004$). The trial-number effect remained null ($b = 0.28$, $p = .57$; original $b = 0.02$, $p = .96$). Audio quality therefore contributes meaningfully to response-quality judgments but does not account for the order or length effects on which our cross-modal results rest.

\section{Additional Synthetic Ratings and Analysis}\label{assitional-synthetic}
We re-run analysis on AI--human agreement on additional synthetic ratings to validate the generalizability of these findings. AI--human agreement is higher for audio models than text models across all three models from differing model families, as seen in Table~\ref{tab:r1}. For rating-gap differentiation, while GPT-4o performs best, all models' differences are predictive of human agreement. When comparing PRISM ratings (human raters interacting with the model) to synthetic ratings, all three judges rate above PRISM, with larger deviations for audio than for text (Table~\ref{tab:r3}). This mirrors the human raters, whose text ratings closely match PRISM while their audio ratings run higher.

\begin{table}[h!]
\centering
\caption{AI--human agreement and rating error.}
\label{tab:r1}
\begin{tabular}{l cc cc cc}
\toprule
 & \multicolumn{2}{c}{$r$} & \multicolumn{2}{c}{MAE} & & \\
\cmidrule(lr){2-3} \cmidrule(lr){4-5}
Rater & Audio & Text & Audio & Text & $\bar{d}$ & $p$ \\
\midrule
GPT-4o              & $.698$ & $.693$ & 12.27 & 14.29 & 2.02 & $.035$ \\
Gemini 3.6 Flash    & $.707$ & $.475$ & 13.59 & 20.05 & 6.46 & $<.001$ \\
Qwen 3.5 Omni Flash & $.723$ & $.625$ & 12.34 & 18.47 & 6.12 & $<.001$ \\
\bottomrule
\end{tabular}
\vspace{4pt}
\begin{minipage}{0.95\linewidth}
\footnotesize
\textit{Note.} AI--human error is lower in audio for every rater, and the gap is roughly $3\times$ larger and far more significant for the two newer raters. Audio correlation is nearly constant across raters ($r = .70$--$.72$) while text correlation degrades, i.e., the audio channel is what stabilizes agreement. Note that $r$ is scale-insensitive whereas MAE is not. Analyses on centered MAE are consistent with those reported here.
\end{minipage}
\end{table}

\begin{table}[h!]
\centering
\caption{AI differentiation predicts human agreement.}
\label{tab:r2}
\begin{tabular}{l ccc cc}
\toprule
 & & & & \multicolumn{2}{c}{Spearman} \\
\cmidrule(lr){5-6}
Rater & AI Diff ($B$) & Modality (Text) & AI Diff $\times$ Mod. & Audio & Text \\
\midrule
GPT-4o              & $0.0141^{***}$ & $-0.010$ & $-0.0027$ & $.23^{*}$   & $.38^{***}$ \\
Gemini 3.6 Flash    & $0.0070^{***}$ & $0.079$  & $-0.0018$ & $.40^{***}$ & $.25^{*}$   \\
Qwen 3.5 Omni Flash & $0.0077^{***}$ & $0.021$  & $-0.0022$ & $.32^{**}$  & $.29^{**}$  \\
\bottomrule
\end{tabular}
\vspace{4pt}
\begin{minipage}{0.95\linewidth}
\footnotesize
\textit{Note.} $^{*}p<.05$, $^{**}p<.01$, $^{***}p<.001$. Every claim in Section~\ref{human-ai-alignment} (\textit{AI Difference Predicts Human Agreement}) holds for all three raters: AI differentiation predicts human ICC in both modalities, with no modality main effect and no interaction. The AI-differentiation slope survives refitting with influential observations removed (Cook's $D$; Gemini $B = 0.0052$; Qwen $B = 0.0083$, both $p < .001$; the Qwen interaction reaches $p = .012$ in that refit but is null in the full model). The smaller slopes for Gemini and Qwen reflect their wider raw AI-difference scale, not a weaker relationship.
\end{minipage}
\end{table}

\begin{table}[h!]
\centering
\caption{PRISM vs.\ AI annotators across synthetic raters.}
\label{tab:r3}
\begin{tabular}{l ccc}
\toprule
Comparison & GPT-4o & Gemini 3.6 Flash & Qwen 3.5 Omni Flash \\
\midrule
PRISM vs.\ AI-Text  & $-10.29^{***}$ & $-3.00$         & $-3.86$         \\
PRISM vs.\ AI-Audio & $-17.62^{***}$ & $-10.93^{***}$ & $-17.36^{***}$ \\
\bottomrule
\end{tabular}
\vspace{4pt}
\begin{minipage}{0.95\linewidth}
\footnotesize
\textit{Note.} Mean signed difference (PRISM $-$ group), raw 1--100 scale; negative values mean the AI judge rated higher than the original PRISM respondent. $^{***}p<.001$; unmarked values are not significant (Gemini $p = .214$, Qwen $p = .107$). Every judge rates above PRISM in both modalities, and the gap is larger in audio than in text for all three. The size of the text offset is rater-dependent (significant only for GPT-4o), reflecting each judge's absolute scale calibration rather than the modality contrast our claims rest on.
\end{minipage}
\end{table}

\clearpage
\section{Voice Ablation}\label{voice-ablation}
\textbf{Design.} We re-ran the main audio arm on 30 of the 100 prompts with three new voices chosen to span expressivity while varying speaker identity and engine. We quantified expressivity as the SD of $f_0$ in semitones (scale-free across pitch registers) and self-consistency as the SD of per-clip mean $f_0$ across the 60 clips per arm. Relative to the baseline (Kokoro af\_heart; drift 0.27 st), the arms are: Qwen3-TTS Serena~\cite{hu2026qwen3}, a second female voice from a different engine at comparable expressivity but a markedly higher register (1.08$\times$, 254 vs.\ 206~Hz; drift 0.70 st); Kokoro am\_michael~\cite{hexgrad_kokoro_82m}, a male voice from the baseline engine (1.47$\times$; drift 0.29 st), which varies expressivity and gender within engine; and Chatterbox~\cite{chatterboxtts2025}, a third engine at the expressivity extreme (1.90$\times$; drift 0.98 st). Expressivity and speaker stability trade off across arms, so we report both (Table~\ref{tab:r8}).

\begin{table}[h]
\centering
\caption{Acoustic characterization of ablation voices.}
\label{tab:r8}
\resizebox{\textwidth}{!}{%
\begin{tabular}{l cc cccc c}
\toprule
Engine / voice & WPM & Artic.\ WPM & $f_0$ mean (Hz) & $f_0$ SD (st) & $f_0$ range (st) & RMS dB SD & $f_0$ SD vs.\ base \\
\midrule
Kokoro af\_heart (baseline) & 157.7 & 187.7 & 205.8 & 2.83 & \phantom{0}9.01 & 6.46 & 1.00$\times$ \\
Kokoro am\_michael          & 143.5 & 174.0 & 114.4 & 4.14 & 15.43 & 6.88 & 1.47$\times$ \\
Chatterbox                  & 200.1 & 218.0 & 147.5 & 5.37 & 16.31 & 6.73 & \textbf{1.90}$\times$ \\
Qwen3-TTS Serena            & 131.2 & 146.9 & 254.4 & 3.04 & 10.30 & 7.98 & 1.08$\times$ \\
\bottomrule
\end{tabular}}
\vspace{4pt}
\begin{minipage}{\linewidth}
\footnotesize
\textit{Note.} A manipulation check for Table~\ref{tab:r7} rather than a replication: it establishes that the four arms differ acoustically in the intended way. Measured over all 60 rendered clips per voice in the ablation (the same 30 prompts as Table~\ref{tab:r7}). $f_0$ SD and range are in semitones, so a higher-pitched voice is not credited with more variability for being higher.
\end{minipage}
\end{table}

\textbf{Content verification.} Because the manipulation holds content fixed, every rendered clip was transcribed and each arm's transcript compared against the baseline transcript of the same text, with human verification of any deviations.

\textbf{Prompt selection.} Prompts were stratified into quintiles of absolute response-length difference (6 per quintile) and, within quintile, selected at evenly spaced positions along the cross-modal disagreement ordering. We avoided selecting on the tails of either variable, which would have guaranteed regression to the mean on re-measurement; cross-modal disagreement was verified after selection. The set preserves the corpus distribution (length difference $Mdn = 122.5$, range 0--1185 vs.\ corpus 103, 0--1185; disagreement $Mdn = 14.0$ vs.\ 13.0), and its projected trial duration (135~s) matches the main study (136~s).

\textbf{Participants.} 48 participants completed all 20 trials, of whom 43 passed the attention check; 5 partial sessions also passed and were retained per the main-study protocol, yielding 48 analyzed raters at 8--12 per clip. Cells ran 33--44 minutes at 74--89\% completion.

\textbf{Perceived audio quality.} Serena's perceived audio quality was markedly lower than the baseline's ($-18.5$ points, $p < .001$, mixed model on per-trial quality ratings), whereas Chatterbox matched the baseline. Serena is also the arm with the lowest reliability (ICC(2,1) $= .154$, at the text level) and weakest synthetic alignment (Table~\ref{tab:r9}), consistent with a noisier human signal for that voice; the Chatterbox match rules out degraded quality as the source of its widened thresholds.

\begin{table}[h]
\centering
\caption{Voice ablation: modality signatures by TTS voice.}
\label{tab:r7}
\resizebox{\textwidth}{!}{%
\begin{tabular}{ll c ccc c cc cc}
\toprule
 & & & \multicolumn{3}{c}{Threshold} & & \multicolumn{2}{c}{Slope $b$} & \multicolumn{2}{c}{ICC} \\
\cmidrule(lr){4-6} \cmidrule(lr){8-9} \cmidrule(lr){10-11}
Condition & Engine / voice & $n$ & $M$ & 95\% CI & $\mathit{Mdn}$ & $r_{rb}$ vs.\ text & Recency & Length & (2,1) & (2,$k$) \\
\midrule
Text (reference) & ---                 & 46 & 41.4 & [37.3, 45.6] & 35.0 & ---           & $-5.07$ & $5.32$ & .147 & .610 \\
Audio (baseline) & Kokoro af\_heart    & 46 & 29.3 & [25.7, 32.9] & 21.0 & $.262^{***}$  & $-2.35$ & $4.73$ & .285 & .784 \\
Ablation A1      & Kokoro am\_michael  & 15 & 28.1 & [24.7, 31.5] & 22.0 & $.280^{***}$  & $-1.47$ & $1.83$ & .226 & .744 \\
Ablation A2      & Chatterbox          & 16 & 36.7 & [33.0, 40.5] & 27.0 & $.091$        & $-3.27$ & $6.68$ & .266 & .773 \\
Ablation A3      & Qwen3-TTS Serena    & 17 & 26.1 & [23.1, 29.1] & 16.0 & $.333^{***}$  & $-1.16$ & $2.99$ & .154 & .636 \\
\bottomrule
\end{tabular}}
\vspace{4pt}
\begin{minipage}{\linewidth}
\footnotesize
\textit{Note.} $^{***}p<.001$; Chatterbox $p=.124$. Ablation A holds response text fixed and varies only the rendering voice. Reference ICCs fall below the published values (.312/.806 audio, .241/.738 text) because they are recomputed on 30 rather than 100 prompts; all conditions here are affected equally. Published 100-prompt values, for orientation: text $M=41.7$ [39.5, 43.9], $\mathit{Mdn}=34.0$; audio $M=27.9$ [26.1, 29.8], $\mathit{Mdn}=20.0$, $r_{rb}=.303$.
\end{minipage}
\end{table}

\begin{table}[h]
\centering
\caption{Equivalence tests (TOST) between voice arms.}
\label{tab:r10}
\resizebox{\textwidth}{!}{%
\begin{tabular}{l ccc ccc l}
\toprule
 & \multicolumn{3}{c}{Raw} & \multicolumn{3}{c}{Scaled} & \\
\cmidrule(lr){2-4} \cmidrule(lr){5-7}
Contrast & Est. & 90\% CI & TOST $p$ & Est. & 90\% CI & TOST $p$ & Verdict \\
\midrule
am\_michael $-$ audio baseline & $-1.40$ & [$-8.51$, 5.72]  & .058 & $+0.048$ & [$-0.096$, 0.192] & \textbf{.014} & raw: not established; scaled: equivalent \\
Serena $-$ audio baseline      & $-1.50$ & [$-8.46$, 5.47]  & .058 & $+0.066$ & [$-0.073$, 0.205] & \textbf{.014} & raw: not established; scaled: equivalent \\
am\_michael $-$ Serena         & $+0.10$ & [$-8.15$, 8.36]  & .058 & $-0.018$ & [$-0.165$, 0.129] & \textbf{.008} & raw: not established; scaled: equivalent \\
Chatterbox $-$ audio baseline  & $+6.86$ & [$-0.29$, 14.01] & .160 & $+0.225$ & [0.081, 0.369]    & .237          & not established \\
Chatterbox $-$ text            & $-4.31$ & [$-11.52$, 2.91] & .118 & $-0.062$ & [$-0.208$, 0.084] & \textbf{.014} & raw: not established; scaled: equivalent \\
\bottomrule
\end{tabular}}
\vspace{4pt}
\begin{minipage}{\linewidth}
\footnotesize
\textit{Note.} Two one-sided tests on the same models as Tables~\ref{tab:r14}--\ref{tab:r15}. The equivalence bound is set at the size of the audio--text effect in this ablation ($\delta = 11.17$ raw rating points, $\delta = 0.287$ scaled), so ``equivalent'' means the arms differ by less than the modality effect the paper reports. TOST $p$-values are Holm-adjusted across the five pre-specified contrasts within each outcome; $p < .05$ supports equivalence, and a non-significant TOST leaves the comparison undetermined in both directions.
\end{minipage}
\end{table}

\begin{table}[h]
\centering
\caption{Synthetic-rating alignment by TTS voice.}
\label{tab:r9}
\small
\begin{tabular}{ll cccc}
\toprule
Voice & Rater & $r$ & $\rho$ & MAE & Signed \\
\midrule
Kokoro af\_heart (baseline) & GPT-4o-Audio\textsuperscript{a} & .682 & .496 & 13.77 & $+11.10$ \\
Kokoro af\_heart (baseline) & Gemini 3.6 Flash     & .719 & .613 & 16.28 & $+3.49$  \\
Kokoro af\_heart (baseline) & Qwen 3.5 Omni Flash  & .663 & .409 & 13.88 & $+11.78$ \\
Kokoro am\_michael          & Gemini 3.6 Flash     & .529 & .521 & 17.85 & $+2.88$  \\
Kokoro am\_michael          & Qwen 3.5 Omni Flash  & .504 & .360 & 15.89 & $+10.99$ \\
Chatterbox                  & Gemini 3.6 Flash     & .567 & .384 & 16.87 & $+1.11$  \\
Chatterbox                  & Qwen 3.5 Omni Flash  & .626 & .480 & 13.98 & $+9.23$  \\
Qwen3-TTS Serena            & Gemini 3.6 Flash     & .596 & .400 & 20.45 & $+9.52$  \\
Qwen3-TTS Serena            & Qwen 3.5 Omni Flash  & .425 & .149 & 20.41 & $+15.64$ \\
\bottomrule
\end{tabular}
\par\smallskip
\begin{minipage}{0.95\linewidth}
\footnotesize
\textit{Note.} Per-clip mean synthetic rating against per-clip mean human rating over the 60 clips of the 30 Ablation-A prompts. Signed $>0$ = synthetic rater above human. \textsuperscript{a}gpt-4o-audio-preview-2024-12-17, as in the main study. Alignment is strongest on the baseline voice and weakest for the highest-pitch voice (Serena), but at 30 prompts individual voice-vs.-baseline differences are not resolvable under a prompt-clustered bootstrap.
\end{minipage}
\end{table}

\section{Audio-Context Ablation}\label{full-audio-ablation}
\textbf{Design.} Arm B1 synthesized the conversation history in a single voice (af\_heart) for both speakers, with a visual indicator of the current speaker during playback. Arm B2 rendered user turns in am\_michael and assistant turns in af\_heart. B1 is less ecologically valid while B2 introduces a vocal-identity confound; comparing them directly separates the contribution of speaker differentiation from that of hearing the history at all. Rated responses remained TTS-rendered in af\_heart in both arms.

\textbf{Prompt selection.} Of the 100 prompts, 64 had conversation history and 62 passed the cross-arm content check. To avoid fatigue confounds we capped projected trial duration at 270.6~s (the 75th percentile of the eligible pool; 46 prompts), yielding a mean of 181~s and $\sim$15 trials per 47-minute session. Remaining prompts were allocated proportionally by turn count and selected at evenly spaced positions along the length-difference ordering. The set matched the eligible pool on length difference ($Mdn = 70.5$ characters, range 0--598) and cross-modal disagreement ($Mdn = 15.9$ vs.\ 14.4); 15 prompts overlap with Ablation A.

\textbf{Participants.} B1: 22 analyzed raters (2 partial) at 9--12 per clip; B2: 23 (1 partial) at 9--14. Median session duration was 46.5 (B1) and 47.8 (B2) minutes against a design target of $\approx$47, with completion rates of 72\% (23/32) and 80\% (24/30).

\textbf{B1 reliability anomaly.} B1's ICC(2,1) of $.032$ warrants scrutiny because the artifact runs in the direction of our headline result: raters who do not differentiate stimuli produce near-zero within-pair gaps, mechanically narrowing the arm's threshold. The low ICC does not reflect rater disagreement: stimulus variance falls to 11.9 against the audio baseline's 108.6 while rater variance stays in range (Table~\ref{tab:r13}). It traces to a non-differentiating subgroup: 10/22 B1 raters used a rating SD under 10 points (45\%), versus 9\% in B2, which shared trial length, task structure, and recruitment window, and 13--15\% in the reference arms on the same prompts. Splitting B1 at the median trial duration (a criterion independent of the ratings), the faster half yields a singular fit with zero stimulus variance, while the slower half yields a threshold of 19.6, matching the audio baseline's 19.0. The narrowing therefore does not depend on the non-differentiating raters. We present this as a post-hoc sensitivity check; from a single cell we cannot tell whether single-voice presentation induced disengagement or B1 drew an unlucky sample, though B2 showing no such subgroup is suggestive.

\textbf{History depth and fatigue.} The audio-context effect does not detectably scale with history depth: adding turn count $\times$ arm to the threshold model leaves the interaction null (LRT $\chi^2(3) = 1.70$, $p = .64$; B1 $\times$ turns $b = +0.16$, 95\% CI $[-3.19, 3.51]$; B2 $\times$ turns $b = -1.04$, $[-4.30, 2.22]$, raw points per SD of turn count); history character length behaves the same (omnibus $p = .067$, with the marginal term in the text reference arm). A trial-position $\times$ condition term shows no differential fatigue in either ablation (LRT $p = .079$ in B, $p = .35$ in A; the only nominally significant position slope is in the text reference).

\begin{table}[h]
\centering
\caption{Audio-context ablation: modality signatures.}
\label{tab:r11}
\resizebox{\textwidth}{!}{%
\begin{tabular}{lll c cccc c cc cc}
\toprule
 & & & & \multicolumn{4}{c}{Threshold} & & \multicolumn{2}{c}{Slope $b$} & \multicolumn{2}{c}{ICC} \\
\cmidrule(lr){5-8} \cmidrule(lr){10-11} \cmidrule(lr){12-13}
Condition & History & Response & $n$ & $M$ & 95\% CI & $\mathit{Mdn}$ & Scaled & $r_{rb}$ vs.\ text & Recency & Length & (2,1) & (2,$k$) \\
\midrule
Text (reference) & text              & text  & 46 & 31.9 & [28.1, 35.7] & 24.5 & 1.44 & ---          & $-3.69^{\dagger}$ & $4.56^{**}$       & .085 & .450 \\
Audio (baseline) & text              & audio & 46 & 19.0 & [16.4, 21.6] & 14.0 & 1.13 & $.333^{***}$ & $-1.66$           & $3.16^{\dagger}$  & .173 & .657 \\
Ablation B1      & audio, one voice  & audio & 22 & 16.5 & [14.3, 18.7] & 11.0 & 1.12 & $.419^{***}$ & $-2.49^{\dagger}$ & $0.64$            & .032 & .250 \\
Ablation B2      & audio, two voices & audio & 23 & 29.1 & [26.4, 31.9] & 25.0 & 1.33 & $.032$       & $+1.64$           & $1.29$            & .069 & .454 \\
\bottomrule
\end{tabular}}
\vspace{4pt}
\begin{minipage}{\linewidth}
\footnotesize
\textit{Note.} $^{***}p<.001$, $^{**}p<.01$, $^{\dagger}p<.10$; B2 threshold $p=.577$, all other coefficients $p>.25$. Ablation B holds prompt text and response text fixed and varies the modality of the conversation history. Reference conditions are recomputed on the same 30 Ablation-B prompts, so they differ from the values in Table~\ref{tab:r7}, which use the 30 Ablation-A prompts.
\end{minipage}
\end{table}

\begin{table}[h]
\centering
\caption{Equivalence tests (TOST) between audio-context arms.}
\label{tab:r10b}
\resizebox{\textwidth}{!}{%
\begin{tabular}{l ccc ccc l}
\toprule
 & \multicolumn{3}{c}{Raw} & \multicolumn{3}{c}{Scaled} & \\
\cmidrule(lr){2-4} \cmidrule(lr){5-7}
Contrast & Est. & 90\% CI & TOST $p$ & Est. & 90\% CI & TOST $p$ & Verdict \\
\midrule
B1 $-$ audio baseline & $-3.36$  & [$-9.66$, 2.94]    & \textbf{.014} & $-0.028$ & [$-0.192$, 0.136]  & \textbf{.024} & equivalent \\
B2 $-$ audio baseline & $+9.55$  & [3.43, 15.67]      & .325          & $+0.193$ & [0.035, 0.352]     & .481          & not established \\
B2 $-$ B1             & $+12.91$ & [5.91, 19.91]      & .772          & $+0.221$ & [0.056, 0.387]     & .500          & not established \\
B1 $-$ text           & $-17.49$ & [$-23.90$, $-11.07$] & .806        & $-0.317$ & [$-0.484$, $-0.149$] & .610        & not established \\
B2 $-$ text           & $-4.58$  & [$-10.82$, 1.66]   & \textbf{.025} & $-0.095$ & [$-0.258$, 0.067]  & .101          & raw: equivalent; scaled: not established \\
\bottomrule
\end{tabular}}
\vspace{4pt}
\begin{minipage}{\linewidth}
\footnotesize
\textit{Note.} Same construction as Table~\ref{tab:r10}, applied to the Ablation B models, with the bound set at this ablation's audio--text effect ($\delta = 14.13$ raw, $\delta = 0.289$ scaled). Holm-adjusted across the five pre-specified contrasts within each outcome. B1's equivalence to the audio baseline on both metrics is the affirmative counterpart of the significance tests in Table~\ref{tab:r15}.
\end{minipage}
\end{table}

\section{Cross-Ablation Results}\label{cross-ablation-results}
Tables~\ref{tab:r12}--\ref{tab:r15} pool both ablations: winner-level agreement against a split-half noise floor, rating-scale usage and variance components, and the mixed models underlying the contrasts reported in Section~\ref{robustness}.

\begin{table}[h]
\centering
\caption{Winner agreement against the text condition, with a within-condition noise floor.}
\label{tab:r12}
\begin{tabular}{l cc cc}
\toprule
 & \multicolumn{2}{c}{Agreement with} & \multicolumn{2}{c}{Split-half floor} \\
\cmidrule(lr){2-3} \cmidrule(lr){4-5}
Condition & Text & B1 & Mean & 95\% CI \\
\midrule
Text (reference) & ---  & .600 & .543 & [.43, .67] \\
Audio (baseline) & .300 & .500 & .564 & [.43, .70] \\
Ablation B1      & .600 & ---  & .548 & [.37, .70] \\
Ablation B2      & .367 & .500 & .555 & [.43, .67] \\
\bottomrule
\end{tabular}
\vspace{4pt}
\begin{minipage}{0.95\linewidth}
\footnotesize
\textit{Note.} Per-prompt winner from clip means. The paper reports audio--text agreement of 53/100 prompts at full corpus size (binomial vs.\ 50\%, $p = .31$), i.e., near chance; the values here are read against that baseline. The floor is the mean agreement between two random halves of the same condition's raters (400 splits), i.e., how often a condition disagrees with itself from sampling noise alone. An agreement is evidence of a genuine difference in winners only if it falls below the floor's interval.
\end{minipage}
\end{table}

\begin{table}[h]
\centering
\caption{Rating-scale usage by condition.}
\label{tab:r13}
\begin{tabular}{l cccc ccc}
\toprule
 & & & & & \multicolumn{3}{c}{Variance components} \\
\cmidrule(lr){6-8}
Condition & Mean rating & Per-rater SD & Raters with SD $<10$ & Tie rate & $\sigma^2_{\mathrm{stim}}$ & $\sigma^2_{\mathrm{rater}}$ & $\sigma^2_{\mathrm{resid}}$ \\
\midrule
Text (reference) & 70.5 & 21.1 & 6/46  & .358 & 59.4  & 170.9 & 470.2 \\
Audio (baseline) & 76.4 & 19.4 & 7/46  & .341 & 108.6 & 163.5 & 356.8 \\
Ablation B1      & 82.5 & 13.5 & 10/22 & .289 & 11.9  & 119.2 & 244.6 \\
Ablation B2      & 75.3 & 20.9 & 2/23  & .254 & 40.0  & 102.8 & 439.4 \\
\bottomrule
\end{tabular}
\end{table}

\begin{table}[h]
\centering
\caption{Estimated marginal means (EMM) for the decision threshold.}
\label{tab:r14}
\begin{tabular}{l ccc ccc}
\toprule
 & \multicolumn{3}{c}{Raw} & \multicolumn{3}{c}{Scaled} \\
\cmidrule(lr){2-4} \cmidrule(lr){5-7}
Condition & EMM & SE & 95\% CI & EMM & SE & 95\% CI \\
\midrule
\multicolumn{7}{l}{\textbf{Ablation A (voice)}} \\
Text                              & 39.03 & 3.19 & [32.72, 45.35] & 1.395 & .093 & [1.208, 1.581] \\
Audio baseline (Kokoro af\_heart) & 27.86 & 3.14 & [21.65, 34.08] & 1.108 & .093 & [0.923, 1.293] \\
Kokoro am\_michael                & 26.47 & 4.11 & [18.34, 34.60] & 1.156 & .097 & [0.963, 1.349] \\
Chatterbox                        & 34.73 & 4.13 & [26.56, 42.90] & 1.332 & .097 & [1.139, 1.526] \\
Qwen3-TTS Serena                  & 26.37 & 4.01 & [18.43, 34.31] & 1.174 & .094 & [0.986, 1.362] \\
\addlinespace
\multicolumn{7}{l}{\textbf{Ablation B (context)}} \\
Text                              & 32.84 & 2.58 & [27.75, 37.93] & 1.409 & .082 & [1.247, 1.570] \\
Audio baseline (Kokoro af\_heart) & 18.71 & 2.47 & [13.83, 23.58] & 1.120 & .079 & [0.964, 1.277] \\
B1 one-voice context              & 15.35 & 3.20 & [9.01, 21.69]  & 1.092 & .083 & [0.926, 1.258] \\
B2 two-voice context              & 28.26 & 3.07 & [22.17, 34.34] & 1.313 & .079 & [1.155, 1.471] \\
\bottomrule
\end{tabular}
\vspace{4pt}
\begin{minipage}{0.95\linewidth}
\footnotesize
\textit{Note.} From one model per ablation and outcome, \texttt{gap $\sim$ Arm + (1|PromptID) + (1|ParticipantID)}, fit on non-tie trials. Prompt is crossed with condition (every arm covers the same 30 prompts); rater is nested within condition, since assignment is between-subjects. \textit{Raw} is the paper's definition, in rating points; \textit{scaled} divides each trial's gap by that rater's own rating SD, so a rater confined to a narrow band is not credited with a narrow threshold. Means are model-adjusted and so differ slightly from the raw values in Tables~\ref{tab:r7} and~\ref{tab:r11}. Satterthwaite degrees of freedom. 
\end{minipage}
\end{table}

\begin{table}[h]
\centering
\caption{Direct pairwise contrasts between conditions.}
\label{tab:r15}
\resizebox{\textwidth}{!}{%
\begin{tabular}{l cccc cccc}
\toprule
 & \multicolumn{4}{c}{Raw} & \multicolumn{4}{c}{Scaled} \\
\cmidrule(lr){2-5} \cmidrule(lr){6-9}
Contrast & Est. & 95\% CI (Tukey) & $p$ adj & $p$ unadj & Est. & 95\% CI (Tukey) & $p$ adj & $p$ unadj \\
\midrule
\multicolumn{9}{l}{\textbf{Ablation A (voice)}} \\
Text $-$ Audio baseline       & $+11.17$ & [1.76, 20.58]     & \textbf{.011} & .001 & $+0.287$ & [0.059, 0.516]    & \textbf{.006} & .001 \\
Text $-$ am\_michael          & $+12.56$ & [0.56, 24.57]     & \textbf{.036} & .004 & $+0.239$ & [$-0.004$, 0.482] & .056          & .007 \\
Text $-$ Serena               & $+12.66$ & [0.90, 24.43]     & \textbf{.028} & .003 & $+0.221$ & [$-0.014$, 0.456] & .077          & .010 \\
Text $-$ Chatterbox           & $+4.31$  & [$-7.76$, 16.37]  & .859          & .324 & $+0.062$ & [$-0.181$, 0.306] & .954          & .480 \\
Chatterbox $-$ Audio baseline & $+6.86$  & [$-5.09$, 18.81]  & .505          & .114 & $+0.225$ & [$-0.016$, 0.465] & .079          & .011 \\
Chatterbox $-$ am\_michael    & $+8.26$  & [$-5.82$, 22.34]  & .481          & .106 & $+0.177$ & [$-0.079$, 0.432] & .308          & .057 \\
Chatterbox $-$ Serena         & $+8.36$  & [$-5.51$, 22.22]  & .453          & .097 & $+0.159$ & [$-0.089$, 0.407] & .386          & .077 \\
am\_michael $-$ Serena        & $+0.10$  & [$-13.72$, 13.92] & 1.000         & .984 & $-0.018$ & [$-0.266$, 0.229] & 1.000         & .837 \\
\addlinespace
\multicolumn{9}{l}{\textbf{Ablation B (context)}} \\
Text $-$ Audio baseline       & $+14.13$ & [5.58, 22.68]     & $\mathbf{<.001}$ & $<.001$ & $+0.289$ & [0.036, 0.541]    & \textbf{.018} & .003 \\
Text $-$ B1                   & $+17.49$ & [7.39, 27.58]     & $\mathbf{<.001}$ & $<.001$ & $+0.317$ & [0.053, 0.580]    & \textbf{.011} & .002 \\
Text $-$ B2                   & $+4.58$  & [$-5.24$, 14.40]  & .617             & .226    & $+0.095$ & [$-0.159$, 0.350] & .764          & .331 \\
B2 $-$ B1                     & $+12.91$ & [1.88, 23.94]     & \textbf{.015}    & .003    & $+0.221$ & [$-0.039$, 0.482] & .124          & .029 \\
B2 $-$ Audio baseline         & $+9.55$  & [$-0.08$, 19.18]  & .053             & .011    & $+0.193$ & [$-0.056$, 0.442] & .186          & .046 \\
\bottomrule
\end{tabular}}
\vspace{4pt}
\begin{minipage}{\linewidth}
\footnotesize
\textit{Note.} All pairwise differences from the models in Table~\ref{tab:r14}, so arms are compared to each other directly rather than by whether their separate tests against text reached significance. Tukey-adjusted intervals and $p$-values control familywise error across all pairs; unadjusted $p$-values are also shown, since the arm-vs-arm comparisons were specified before the tests were run. Positive estimates mean the first condition has the wider threshold.
\end{minipage}
\end{table}

\section{Qualitative Analysis}\label{qualitative}
To compare how participants explained their preferences across modalities, we analyzed 658 written justifications (347 audio, 311 text) using TF-IDF with Porter stemming (vocabulary = 223 terms). For each term, we tested whether document frequency differed by modality using $\chi^2$ tests on presence/absence, with Benjamini-Hochberg FDR correction ($q = .05$). Excluding modality-specific terms, audio justifications more frequently mentioned ``user'' ($\chi^2 = 12.8$, $p < .001$) and ``help'' ($\chi^2 = 9.8$, $p = .002$), while text justifications more frequently mentioned ``response'' ($\chi^2 = 14.5$, $p < .001$), ``detail'' ($\chi^2 = 14.5$, $p < .001$), and ``give'' ($\chi^2 = 17.0$, $p < .001$), suggesting audio raters focused more on whether the response served the user's needs while text raters attended more to the depth and specificity of the response content.

\subsection{Maximal cross-modal disagreements}
We ranked all 100 prompts by the absolute difference between the text and audio preference gaps and grouped justifications by modality; the full ranking is released with the data. All three top cases flip the winner and illustrate distinct mechanisms.

\textbf{Verbosity} (``Talk about an interesting fact about fishes.''; $\Delta = 66.9$). Response A gave three elaborated facts, B gave one. Text raters preferred A ($M = 82.2$ vs.\ 47.8); audio raters preferred B ($M = 90.7$ vs.\ 58.2). Text: ``gave interesting facts about fishes as requested, B only gave one.'' Audio: ``Response A went off course by providing 3 facts when the user simply asked for an interesting fact.'' The same verbosity reads as thoroughness in text and as filler in audio.

\textbf{Narration} (``Generate a 6 month professional wrestling story arc.''; $\Delta = 45.5$). Response B is narratively richer but truncates mid-arc and crowns the champion three months in. Text raters penalized it ($M = 44.9$ vs.\ 72.9); audio raters preferred it ($M = 90.1$ vs.\ 72.6): ``B has glitch sounds. B tells a much better story though.'' The truncation is noticed in both modalities but weighted differently; the constraint violation is what listeners miss, since verifying arc structure across 72~s of narration is harder than checking it at a glance. We do not filter truncated generations, as early stopping is ecologically valid for deployed audio systems.

\textbf{Irrelevant prose} (``I may look for something a little lighter to watch.''; $\Delta = 43.5$). Response B appends a hallucinated dialogue continuation to an otherwise reasonable recommendation. Text raters barely penalized it ($M = 68.4$ vs.\ 66.4; ``extra junk,'' ``oddly formatted''); audio raters penalized it severely ($M = 33.7$ vs.\ 75.2; ``just went off the rails''). Formatting artifacts can be skipped in text but not in audio, where they become consequential disruptions.

Descriptively, verifiable content advantages the reader and experiential content the listener; at the current sample this is a hypothesis rather than a result.

\subsection{Prompt-type induction}
To avoid imposing our own categories, we induced a taxonomy over the 100 prompts with TopicGPT~\cite{pham-etal-2024-topicgpt}: iterative topic generation seeded with two format-demonstration topics, Sentence-Transformer merging of near-duplicates at cosine $\geq .5$, then multi-label assignment with supporting quotations and self-correction. Generation converged quickly (8 topics in the first 20 prompts, 1 more across the remaining 80), yielding 11 topics. Assignment was run independently by two model families, which agreed on 75\% of prompts (Cohen's $\kappa = .699$). No coding topic emerged, since PRISM's unguided conversations contain none; creative writing is present but rare ($n = 3$). Descriptively, the largest modality swings fall on open-ended and creative requests (both favored in audio) and the smallest on factual and explanatory ones, but at these cell sizes we treat this as a hypothesis (Table~\ref{tab:r6}).

\begin{table}[h]
\centering
\caption{Modality effect by induced prompt type.}
\label{tab:r6}
\begin{tabular}{l cccc}
\toprule
Prompt type & $n$ & $|$gap$|$ & Signed & Flips \\
\midrule
Open-ended Information Request & 6  & $27.3$ & $-19.3$ & 33\% \\
Creative Writing Request       & 3  & $23.1$ & $-16.5$ & 67\% \\
Follow-up Request              & 12 & $18.1$ & $+4.1$  & 83\% \\
Recommendation Request         & 10 & $17.3$ & $-0.6$  & 40\% \\
Small Talk                     & 4  & $17.0$ & $+12.4$ & 25\% \\
Opinion/Advice Request         & 30 & $14.7$ & $-0.5$  & 60\% \\
Explanation Request            & 7  & $13.2$ & $-0.8$  & 29\% \\
Factual Question               & 25 & $12.8$ & $-2.7$  & 36\% \\
\bottomrule
\end{tabular}
\vspace{4pt}
\begin{minipage}{0.95\linewidth}
\footnotesize
\textit{Note.} Topics with $n \geq 3$. Signed $> 0$ = favoured in text. Procedure in text above. Prompt type does \textbf{not} significantly predict the modality effect. Adding topic to a length-only model does not improve fit ($R^2$: $.053 \rightarrow .129$; $F = 1.43$, $p = .22$). Length asymmetry therefore remains the better predictor, and we report this null directly.
\end{minipage}
\end{table}

\section{Triage Analysis}\label{triage-analysis}
\begin{table}[h]
\centering
\caption{First-hit vs.\ stable ratings by gap band.}
\label{tab:r4}
\begin{tabular}{l cccc}
\toprule
Gap band & $n$ & First-hit mean & Stable mean & \%\,$k{=}1$ \\
\midrule
\multicolumn{5}{l}{\textbf{Audio}} \\
0--1  & 19 & 2.36 & 3.32 & 52 \\
1--2  & 17 & 2.12 & 3.35 & 57 \\
2--4  & 23 & 2.33 & 3.98 & 54 \\
4--9  & 20 & 2.03 & 3.27 & 60 \\
9--82 & 20 & 1.58 & 2.33 & 76 \\
\textit{all} & \textit{99} & \textit{2.09} & \textit{3.27} & \textit{60} \\
\addlinespace
\multicolumn{5}{l}{\textbf{Text}} \\
0--3   & 20 & 2.43 & 3.33 & 49 \\
3--5   & 20 & 2.17 & 3.12 & 56 \\
5--11  & 20 & 2.05 & 2.88 & 59 \\
11--23 & 20 & 1.90 & 2.98 & 64 \\
23--74 & 20 & 1.47 & 1.98 & 78 \\
\textit{all} & \textit{100} & \textit{2.00} & \textit{2.86} & \textit{61} \\
\bottomrule
\end{tabular}
\vspace{4pt}
\begin{minipage}{0.95\linewidth}
\footnotesize
\textit{Note.} Within-modality quintiles. First-hit is the first $k$ matching the golden winner; stable is the smallest $k$ matching at every larger $k$. \%\,$k{=}1$ is the first-hit rate at a single rater ($\sim$50\% is chance). Panel size 8--10; one audio prompt excluded for an exactly tied golden margin. In sum, with a gap larger than 9 in audio we can suggest 2 annotators instead of 3. In text, we would suggest 2 annotators instead of 3 for a gap larger than 23.
\end{minipage}
\end{table}

\clearpage
\section{Participant Demographics}
Table~\ref{tab:demographics} reports the demographic composition of the
final analytic sample ($N = 94$) by modality condition. We report these
descriptives both for transparency and to support the robustness check
described below.

\label{app:demographics}
\begin{table}[h]
\caption{Participant Demographics by Condition}
\label{tab:demographics}
\centering
\small
\begin{tabular}{lccc}
\hline
& \textbf{Text} ($n=48$) & \textbf{Audio} ($n=46$) & \textbf{Total} ($N=94$) \\
\hline
\textit{Age (years)} & & & \\
\quad $M$ ($SD$) & 45.00 (13.77) & 40.17 (11.70) & 42.64 (12.96) \\
\quad Median & 44 & 40 & 41 \\
\quad Range & 22--75 & 20--72 & 20--75 \\
\addlinespace
\textit{Gender} & & & \\
\quad Woman & 31 (64.6\%) & 25 (54.3\%) & 56 (59.6\%) \\
\quad Man & 17 (35.4\%) & 20 (43.5\%) & 37 (39.4\%) \\
\quad Agender & 0 (0.0\%) & 1 (2.2\%) & 1 (1.1\%) \\
\addlinespace
\textit{Race} & & & \\
\quad White & 29 (60.4\%) & 30 (65.2\%) & 59 (62.8\%) \\
\quad Black or African American & 7 (14.6\%) & 10 (21.7\%) & 17 (18.1\%) \\
\quad Asian & 7 (14.6\%) & 2 (4.3\%) & 9 (9.6\%) \\
\quad Two or More Races & 3 (6.2\%) & 3 (6.5\%) & 6 (6.4\%) \\
\quad American Indian/Alaska Native & 2 (4.2\%) & 0 (0.0\%) & 2 (2.1\%) \\
\quad Prefer not to say & 0 (0.0\%) & 1 (2.2\%) & 1 (1.1\%) \\
\addlinespace
\textit{Ethnicity} & & & \\
\quad Not Hispanic or Latino & 44 (91.7\%) & 39 (84.8\%) & 83 (88.3\%) \\
\quad Hispanic or Latino & 3 (6.2\%) & 7 (15.2\%) & 10 (10.6\%) \\
\quad Prefer not to say & 1 (2.1\%) & 0 (0.0\%) & 1 (1.1\%) \\
\addlinespace
\textit{English Proficiency} & & & \\
\quad Native speaker & 41 (85.4\%) & 42 (91.3\%) & 83 (88.3\%) \\
\quad Fluent & 6 (12.5\%) & 4 (8.7\%) & 10 (10.6\%) \\
\quad Conversational & 1 (2.1\%) & 0 (0.0\%) & 1 (1.1\%) \\
\addlinespace
\textit{Education} & & & \\
\quad Bachelor's degree & 21 (43.8\%) & 14 (30.4\%) & 35 (37.2\%) \\
\quad Some college, no degree & 7 (14.6\%) & 13 (28.3\%) & 20 (21.3\%) \\
\quad High school or equivalent & 8 (16.7\%) & 8 (17.4\%) & 16 (17.0\%) \\
\quad Master's degree & 6 (12.5\%) & 5 (10.9\%) & 11 (11.7\%) \\
\quad Associate degree & 4 (8.3\%) & 5 (10.9\%) & 9 (9.6\%) \\
\quad Doctorate & 2 (4.2\%) & 1 (2.2\%) & 3 (3.2\%) \\
\addlinespace
\quad Bachelor's+ (\%) & 60.4 & 43.5 & 52.1 \\
\hline
\end{tabular}
\par\smallskip\noindent\footnotesize
\textit{Note.} All participants resided in the United States. Demographics reflect participants who passed the attention check.
\end{table}

Participant demographics were largely comparable across the two modality conditions. Both groups were predominantly White (Text: 60.4\%, Audio: 65.2\%), non-Hispanic (91.7\%, 84.8\%), and native English speakers (85.4\%, 91.3\%). Text-condition participants were somewhat older on average ($M = 45.00$, $SD = 13.77$) than audio-condition participants ($M = 40.17$, $SD = 11.70$), and held bachelor's degrees or higher at a higher rate (60.4\% vs.\ 43.5\%). As a robustness check, we re-fit the omnibus model with education (bachelor's+ vs. below) as an additional fixed effect; the modality effects reported in the main text were unchanged. Because participants were randomly assigned to modality, these differences reflect sampling variability rather than systematic selection.

\newpage
\input{checklist.tex}

\end{document}

%% file: checklist.tex
\section*{NeurIPS Paper Checklist}

\begin{enumerate}

\item {\bf Claims}
    \item[] Question: Do the main claims made in the abstract and introduction accurately reflect the paper's contributions and scope?
    \item[] Answer: \answerYes{} 
    \item[] Justification: This work is a step towards best practices in preferences data collection for both text and audio modalities. We focus on practical guidelines and novel understandings of how current speech generation pipelines may be different than text pipelines and how to collect higher quality data cross-modality. 

\item {\bf Limitations}
    \item[] Question: Does the paper discuss the limitations of the work performed by the authors?
    \item[] Answer: \answerYes{} 
    \item[] Justification: We explicitly state that our findings related to speech are limited in that they are conducted on a single TTS voice and further works should investigate this further. Additionally, our synthetic data reflects GPT-4o and GPT-4o-Audio-Preview and additional works are needed to generalize these claims. 

\item {\bf Theory assumptions and proofs}
    \item[] Question: For each theoretical result, does the paper provide the full set of assumptions and a complete (and correct) proof?
    \item[] Answer: \answerNA{} 
    \item[] Justification: This papers focus is on practical considerations of data collection for speech and text preferences.

    \item {\bf Experimental result reproducibility}
    \item[] Question: Does the paper fully disclose all the information needed to reproduce the main experimental results of the paper to the extent that it affects the main claims and/or conclusions of the paper (regardless of whether the code and data are provided or not)?
    \item[] Answer: \answerYes{} 
    \item[] Justification: We fully outline our methods, in detail, such that others can convert the PRISM data to speech using kokoro, qwen, or chatterbox and running a similar study using our code or a similar interface to the image presented in the paper.

\item {\bf Open access to data and code}
    \item[] Question: Does the paper provide open access to the data and code, with sufficient instructions to faithfully reproduce the main experimental results, as described in supplemental material?
    \item[] Answer: \answerYes{} 
    \item[] Justification: We make both our data and code available to reproduce our study plus provide instructions on how to set up similar studies.

\item {\bf Experimental setting/details}
    \item[] Question: Does the paper specify all the training and test details (e.g., data splits, hyperparameters, how they were chosen, type of optimizer) necessary to understand the results?
    \item[] Answer: \answerNA{} 
    \item[] Justification: Models were not trained in this study, but analysis and data collection methods are clearly articulated.

\item {\bf Experiment statistical significance}
    \item[] Question: Does the paper report error bars suitably and correctly defined or other appropriate information about the statistical significance of the experiments?
    \item[] Answer: \answerYes{} 
    \item[] Justification: We report appropriate p-values and confidence intervals where relevant, and make statistical adjustments where relevant (Benjamini-Hochberg FDR correction in qualitative analysis)

\item {\bf Experiments compute resources}
    \item[] Question: For each experiment, does the paper provide sufficient information on the computer resources (type of compute workers, memory, time of execution) needed to reproduce the experiments?
    \item[] Answer: \answerYes{} 
    \item[] Justification: All analyses were run on a standard laptop CPU; no GPU or cluster compute was required. Bootstrap analyses (2,000–5,000 iterations) completed in minutes.
    
\item {\bf Code of ethics}
    \item[] Question: Does the research conducted in the paper conform, in every respect, with the NeurIPS Code of Ethics \url{https://neurips.cc/public/EthicsGuidelines}?
    \item[] Answer: \answerYes{} 
    \item[] Justification: Data is collected and anonymized. Data was identified as minimal risk by our institution IRB.

\item {\bf Broader impacts}
    \item[] Question: Does the paper discuss both potential positive societal impacts and negative societal impacts of the work performed?
    \item[] Answer: \answerYes{} 
    \item[] Justification: Discussed in Appendix~\ref{impacts}; see also Limitations (Section~\ref{limitations}) and
  Appendix~\ref{app:demographics}.
    
\item {\bf Safeguards}
    \item[] Question: Does the paper describe safeguards that have been put in place for responsible release of data or models that have a high risk for misuse (e.g., pre-trained language models, image generators, or scraped datasets)?
    \item[] Answer: \answerYes{} 
    \item[] Justification: Our released assets consist of TTS-converted audio of the publicly available PRISM dataset's unguided conversations, which were screened to remove offensive, discriminatory, or emotionally distressing content. We do not release pretrained models, and the data poses no meaningful misuse or dual-use risk beyond what already exists in the source PRISM corpus.

\item {\bf Licenses for existing assets}
    \item[] Question: Are the creators or original owners of assets (e.g., code, data, models), used in the paper, properly credited and are the license and terms of use explicitly mentioned and properly respected?
    \item[] Answer: \answerYes{} 
    \item[] Justification: We cite all assets used: the PRISM dataset (Kirk et al., 2024), the Kokoro TTS model (hexgrad), Qwen 3 TTS, Chatterbox, TTS-Arena benchmark, GPT-4o and GPT-4o-Audio-Preview (OpenAI APIs), Gemini 3.6 Flash, Qwen 3.5 Omni Flash and Prolific for participant recruitment. We use these assets in accordance with their respective licenses and terms of service. 

\item {\bf New assets}
    \item[] Question: Are new assets introduced in the paper well documented and is the documentation provided alongside the assets?
    \item[] Answer: \answerYes{} 
    \item[] Justification: We release the 3,113 TTS-converted audio conversations, data annotations, interface code, and our analysis code with documentation. The dataset is derived from the publicly released PRISM corpus, whose participants consented to public release of their conversations. Documentation of selection, removals, and replacements is provided alongside the released data.

\item {\bf Crowdsourcing and research with human subjects}
    \item[] Question: For crowdsourcing experiments and research with human subjects, does the paper include the full text of instructions given to participants and screenshots, if applicable, as well as details about compensation (if any)? 
    \item[] Answer: \answerYes{} 
    \item[] Justification: We recruited 106 participants from Prolific and compensated them at \$12/hour, above the US federal minimum wage. The study interfaces, including all instructions and rating instruments shown to participants, are presented in Appendix Figures~\ref{fig:study_interface_audio} and~\ref{fig:study_interface_text}. The demographic questionnaire, attention check procedure, exclusion criteria, and task structure are described in Section~\ref{app:demographics}, with full participant demographics reported in Appendix Table~\ref{tab:demographics}.

\item {\bf Institutional review board (IRB) approvals or equivalent for research with human subjects}
    \item[] Question: Does the paper describe potential risks incurred by study participants, whether such risks were disclosed to the subjects, and whether Institutional Review Board (IRB) approvals (or an equivalent approval/review based on the requirements of your country or institution) were obtained?
    \item[] Answer: \answerYes{} 
    \item[] Justification: The study was deemed minimal risk by our institutions review board. We report this explicitly in the paper.

\item {\bf Declaration of LLM usage}
    \item[] Question: Does the paper describe the usage of LLMs if it is an important, original, or non-standard component of the core methods in this research? Note that if the LLM is used only for writing, editing, or formatting purposes and does \emph{not} impact the core methodology, scientific rigor, or originality of the research, declaration is not required.
    \item[] Answer: \answerYes{} 
    \item[] Justification: While LLMs were used to assist with manuscript copy editing, the authors retain full responsibility for study design, analysis, and full manuscript content.

\end{enumerate}